\documentclass{jfm}

\usepackage{graphicx}
\usepackage{dcolumn}
\usepackage{bm}
\usepackage[utf8]{inputenc}
\usepackage{subfig}
\usepackage{graphicx}
\usepackage{ragged2e}
\usepackage{amsmath}
\usepackage{mathtools}
\usepackage{caption}
\usepackage{color}
\usepackage[most]{tcolorbox}
\usepackage{xcolor}
\usepackage{float}
\usepackage{bigints}
\usepackage[export]{adjustbox}
\usepackage{tabularx}
\usepackage{epstopdf, epsfig}
\usepackage[toc,page]{appendix}
\usepackage{nomencl}
\usepackage{mathtools,amssymb,lipsum}
\usepackage{newtxtext}
\usepackage{newtxmath}
\usepackage{natbib}
\usepackage{hyperref}
\hypersetup{
    colorlinks = true,
    urlcolor   = blue,
    citecolor  = black,
}

\newcommand{\RomanNumeralCaps}[1]
\linenumbers

\title{Jet from a very large, surface-gravity wave}

\author{Lohit Kayal\aff{1},
 \and Ratul Dasgupta\aff{1}\corresp{\email{dasgupta.ratul@gmail.com}}}

\affiliation{\aff{1}Chemical Engineering, Indian Institute of Technology, Bombay, India}

\begin{document}
\maketitle
\newcommand{\mj}{{\mathrm{J}}}
\begin{abstract}
We demonstrate that gravity acting alone at large length scales, can produce a jet from a large amplitude, axisymmetric surface deformation imposed on a quiescent, deep pool of liquid. Mechanistically, the jet owes it origin to the focussing of a concentric, surface wave towards the axis of symmetry, quite analogous to such focussing of capillary waves and resultant jet formation, observed during bubble collapse at small scales. A weakly non-linear theory based on the method of multiple scales and the potential flow limit, is presented for a modal (single mode) initial condition representing the solution to the primary Cauchy-Poisson problem. A pair of novel, coupled, amplitude equations are derived governing the modulation of the primary mode. For moderate values of the perturbation parameter $\epsilon$ (a measure of the initial perturbation amplitude), our second order theory captures the overshoot (incipient jet) at the axis of symmetry quite well, demonstrating good agreement with numerical simulation of the incompressible, Euler's equation with gravity \citep{basilisk} and no surface tension. Expectedly, our theory becomes inaccurate  as $\epsilon$ approaches unity. In this strongly nonlinear regime, slender jets form with surface accelerations exceeding gravity by three orders of magnitude. In this inertial regime, the jets observed in our simulations show excellent agreement with the inertial, self-similar, analytical solution by \cite{longuet1983bubbles}. The physical mechanism of axisymmetric jet formation is explained based on mass conservation arguments. We demonstrate that the underlying wave focussing mechanism, may be understood in terms of radially inward motion of nodal points of a linearised, axisymmetric, standing wave. This explanation rationalizes the ubiquitous observation of such jets accompanying cavity collapse phenomena, spanning length scales from microns to several meters. This study, to our knowledge, is the first analytical demonstration of jets created purely under gravity. They represent a more spectacular, axisymmetric analogue of the two dimensional ones observed by \cite{longuet2001vertical_2}.
\end{abstract}


\section{Introduction}
An interesting theoretical study by \cite{miles1968cauchy} begins with reference to an insightful quote from \cite{van1968tsunamis} (italics by us): ``....\textit{the concentric circulate ridges that surround the crater Orientale at lat. $20^{\circ}$S and long. $95^{\circ}$W on the moon may have been initiated as gravity waves on a viscous liquid under the impact of a meteorite}.'' \cite{miles1968cauchy} subsequently presents a novel analytical solution to the viscous, linear, Cauchy-Poisson problem in cylindrical, axisymmetric coordinates motivated by the need to validate Van-Dorn's hypothesis. 
The inviscid, irrotational, linear, Cauchy-Poisson \textit{initial value problem} (IVP) and its solution \citep{cauchy1816theorie,poisson1815memoire} for a general class of initial conditions was reported more than one hundred and fifty years before the viscous extension to the same by \cite{miles1968cauchy} for a localised surface perturbation. It represented possibly, the first instance when integral superposition was employed to answer the IVP posed by Laplace (page $2$, \cite{craik2004origins}) viz. what is the linear response of a body of liquid with a free surface which has been perturbed in a prescribed manner, either by deforming the surface or via application of an initial impulse distribution along it? Our topic of interest here is the gravity-induced collapse of a large cavity and a resultant jet which may be ejected during such collapse. As will be seen below, the Cauchy-Poisson solution provides an extremely useful linear reference, against which the formation of this jet and its temporal evolution may be compared, particularly in order to elucidiate the role of non-linearity in the process.

Analytical progress in the Cauchy-Poisson (CP hereafter) solution is made via usage of Fourier (Cartesian) or Hankel (cylindrical geometry) transforms. Expectedly linearised CP prediction for the surface response remains in integral form (see \cite{lamb1924hydrodynamics}, art. 238 and expressions $3.2.10$ and $3.2.12$ in \cite{debnath1994nonlinear}). Except for very special choices of initial conditions, these Fourier (Hankel) integrals cannot typically be expressed in closed form. Taking recourse to techniques such as Kelvin's method of stationary phase \citep{lamb1924hydrodynamics} ($\hat{x}\rightarrow\infty,\;\hat{t}\rightarrow\infty$ with $\frac{\hat{x}}{\hat{t}} = \text{constant}$), one may asymptotically evaluate the integrals and accurately predict the far-field response at large time. Such knowledge is of some practical interest e.g. in the prediction of waves due to landslides \citep{law1968water}. On the other hand and as will be seen below, an understanding of the near field response may also be interesting, particularly when the initial perturbation has large amplitude. For example, the ejecta sheet seen in fig. $1$a in \cite{range2022chicxulub}, may be modelled in a first approximation as an infinitely tall spike of cross sectional area $A_0$ (i.e. $\hat{\eta}(\hat{x},0)=A_0\delta(\hat{x})$) and one can ask what surface waves may be produced subsequently, due to this? As explained in \cite{lamb1924hydrodynamics} (section $238$-$241$) for the infinite depth case and also in \cite{pidduck1912wave} (taking compressibility and finite depth effects into account) the surface response $\hat{\eta}\left(\hat{x},\hat{t}\right)$, due to such a delta function initial disturbance evolves self-similarly under a gravitational field as
\begin{eqnarray}
	\frac{\pi\; \hat{x}\;\hat{\eta}(\hat{x},\hat{t})}{A_0} = f\left(\frac{g\hat{t}^2}{2\hat{x}}\right),\label{0}
\end{eqnarray}
where the functional form of $f(\cdot)$ in \ref{0} is given in the form of an infinite series \citep{lamb1924hydrodynamics}. Obviously, the assumption of a delta function initial perturbation within a linear framework is suspect but nevertheless provides a reference against which one may compare nonlinear, large amplitude behaviour as we will do subsequently here. 

In the current study, our interest lies in large amplitude, axisymmetric, surface gravity waves. For an initial surface perturbation and zero initial surface impulse ($\hat{\phi}\left(\hat{r},\hat{z}=0,\hat{t}=0\right)=0$) on a radially unbounded pool of infinite depth, the CP surface response is predicted to be
\begin{eqnarray}
\hat{\eta}(\hat{r},\hat{t})=\int_{0}^{\infty}dk\;k\mathrm{J}_0(k\hat{r})\mathbb{H}\left[\hat{\eta}(\hat{r},0)\right]\cos \left(\sqrt{gk}\hat{t}\right), \label{1}
\end{eqnarray}
where $\mathbb{H}\left[\cdot\right]$ represents the zeroth order Hankel transform and $\mathrm{J}_0(\cdot)$ is the zeroth order Bessel function of the first kind. When the initial surface perturbation, $\hat{\eta}(\hat{r},\hat{t}=0)$ is localised with finite width (i.e. has a characteristic length-scale), the CP integral in eqn. \ref{1} needs to be evaluated numerically and does not lead to self-similar evolution of the interface. It will be seen in the next section from numerical simulations, that when the amplitude of the localised perturbation is sufficiently large, slender jets at the symmetry axis develop, shooting upwards and reaching heights far exceeding the initial perturbation amplitude.  
\begin{figure}
	\centering
	\subfloat[Collapsing cavity and it's jet]{\includegraphics[scale=0.28]{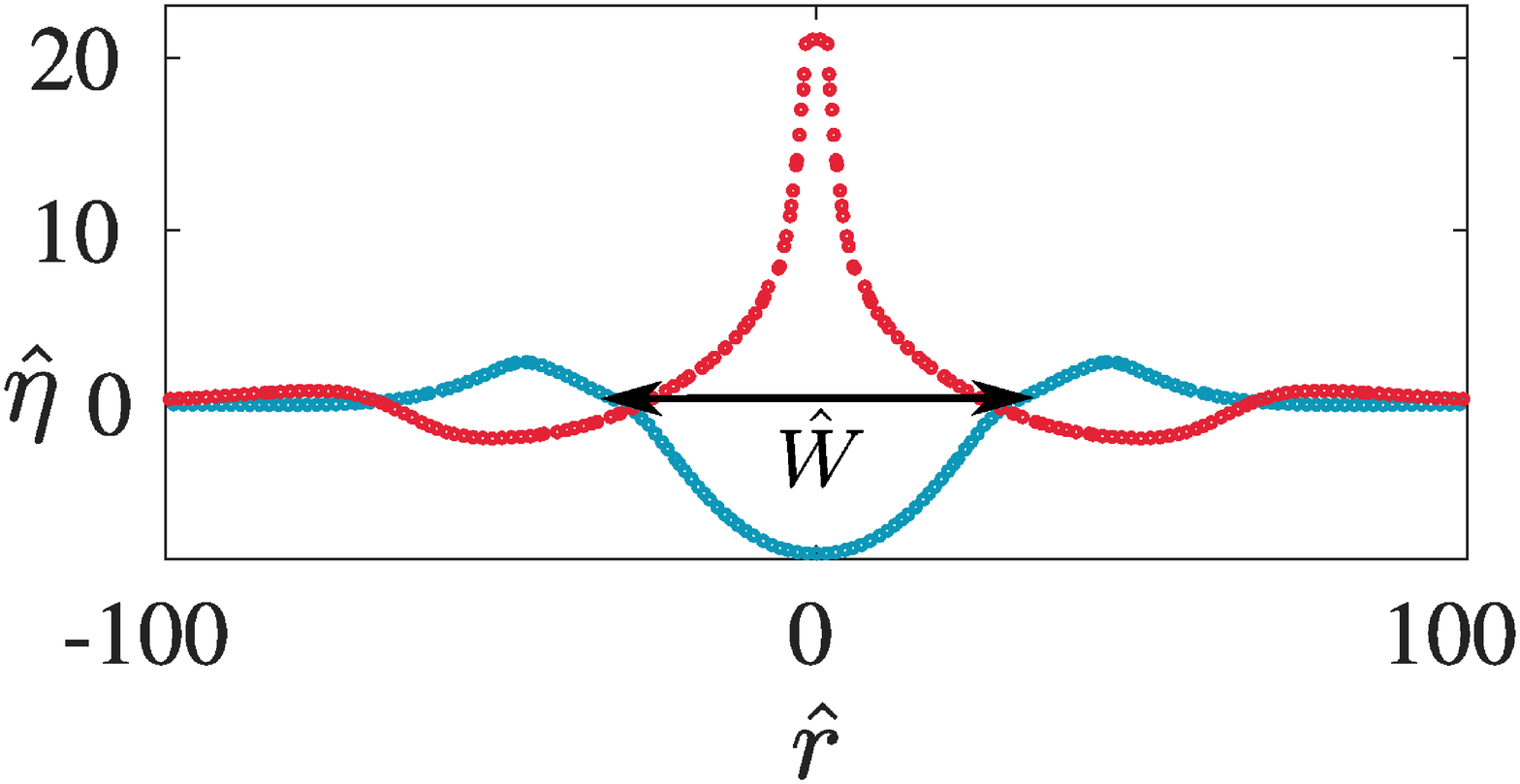}\label{fig_cavity}}
	\subfloat[Collapsing bubble and it's jet]{\includegraphics[scale=0.19]{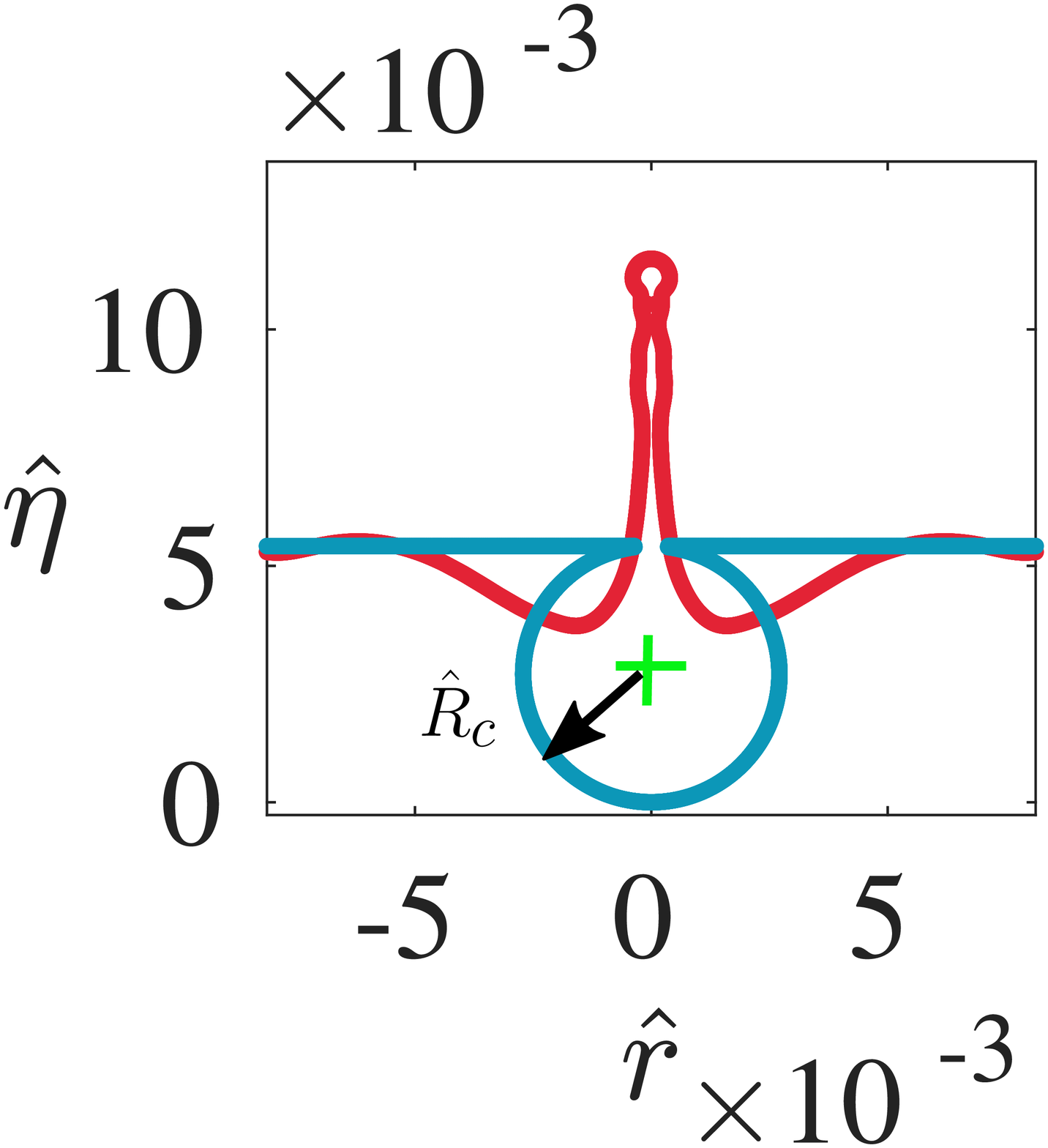}\label{fig_bubble}}
	\caption{Jet formation due to cavity collapse at length-scales separated by approx. five orders of magnitude. Panel (a) A cavity (blue) at an air-water interface generated numerically \citep{basilisk} from an initial hump (see inset of fig. \ref{fig_miles2_a}) of the form $\hat{\eta}(\hat{r},0) = \hat{a}_0\exp\left(-\frac{\hat{r}^2}{\hat{d}^2}\right)\left[1-\left(\frac{\hat{r}}{\hat{d}}\right)^2\right]$ with $\beta \equiv \frac{\hat{a}_0}{\hat{d}} \equiv 0.584$. The cavity width $\hat{W}\approx 62.1$ cms. A jet (red curve) is formed due to the cavity collapse rising sharply upwards at the symmetry axis, $\hat{r}=0$. Panel (b) A tiny bubble (blue) of radius $2.72\times10^{-3}$ cm at the air-water interface whose collapse produces a jet (red curve) ejecting droplets at its tip. Due to the length scale disparity between the panels, the cavity collapse is dictated nearly entirely by gravity in fig. \ref{fig_cavity} and almost entirely by surface tension in \ref{fig_bubble}. The qualitative similarity between the two jets should be noted.} 
\end{figure}
\subsection{Evolution of a cavity}
\begin{figure}
	\centering
	\subfloat[$\hat{t}=0.32$ s]{\includegraphics[scale=0.14]{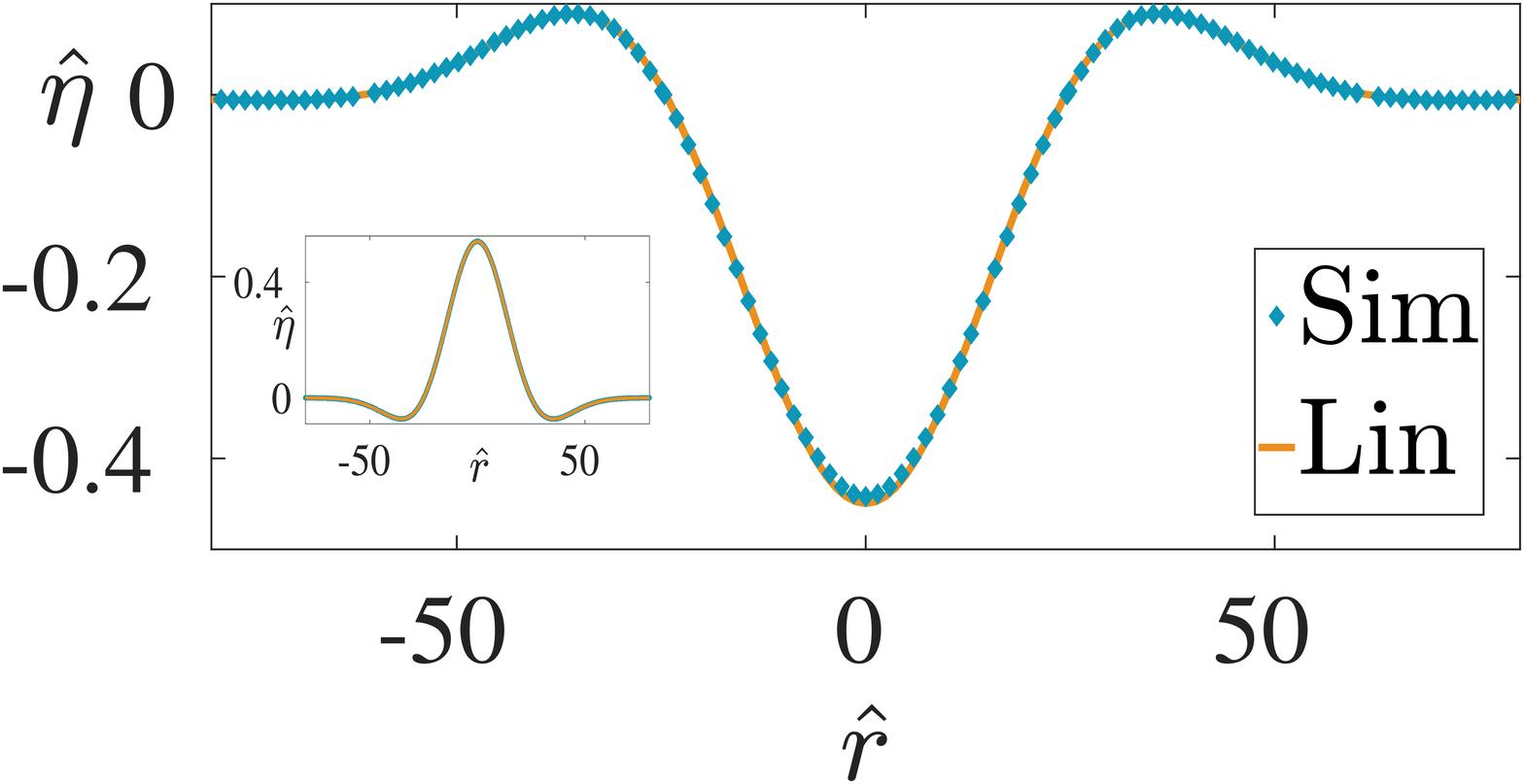}\label{fig_miles1_a}}
	\subfloat[$\hat{t}=0.46$ s]{\includegraphics[scale=0.14]{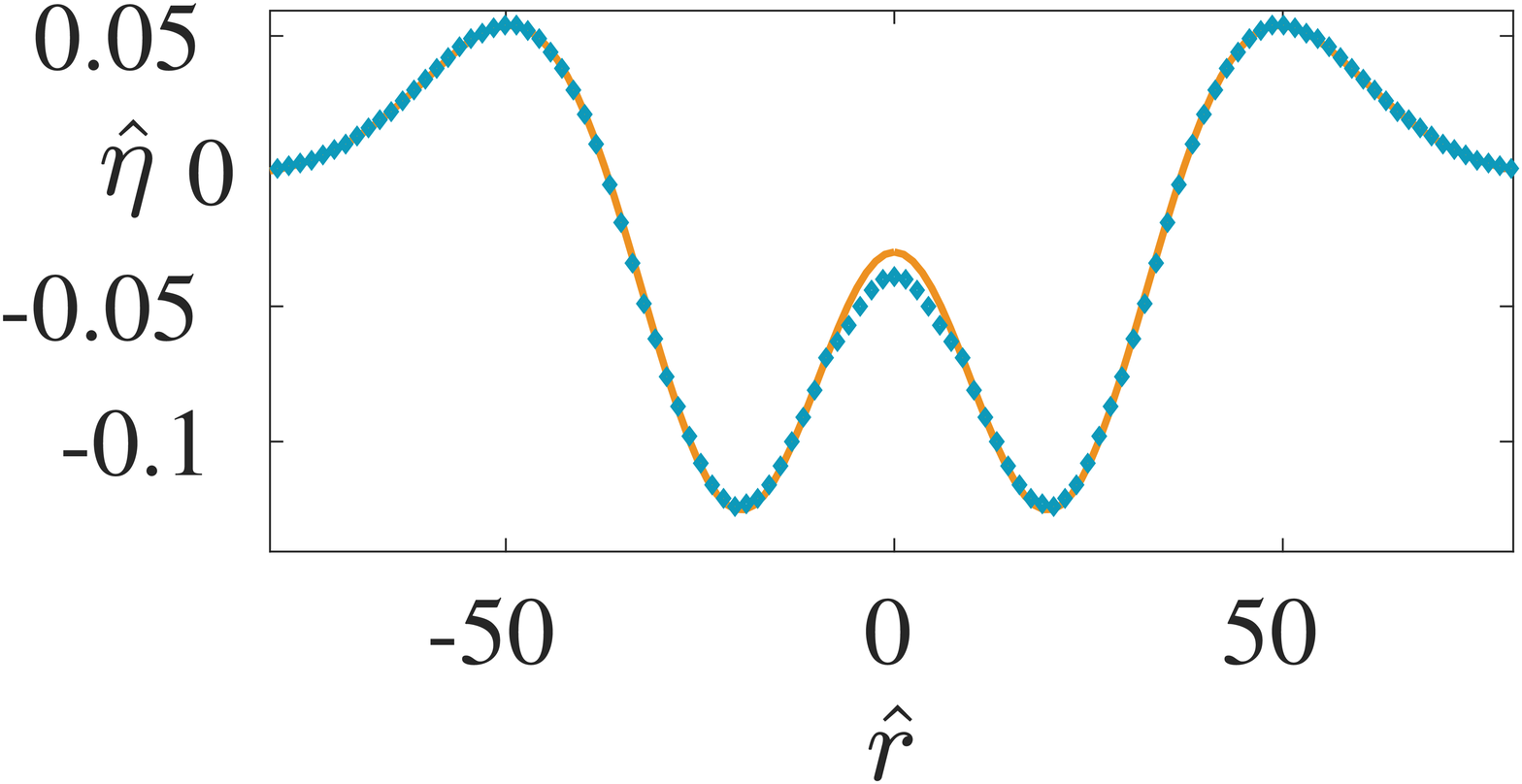}\label{fig_miles1_b}}\\
	\subfloat[$\hat{t}=0.595$ s]{\includegraphics[scale=0.14]{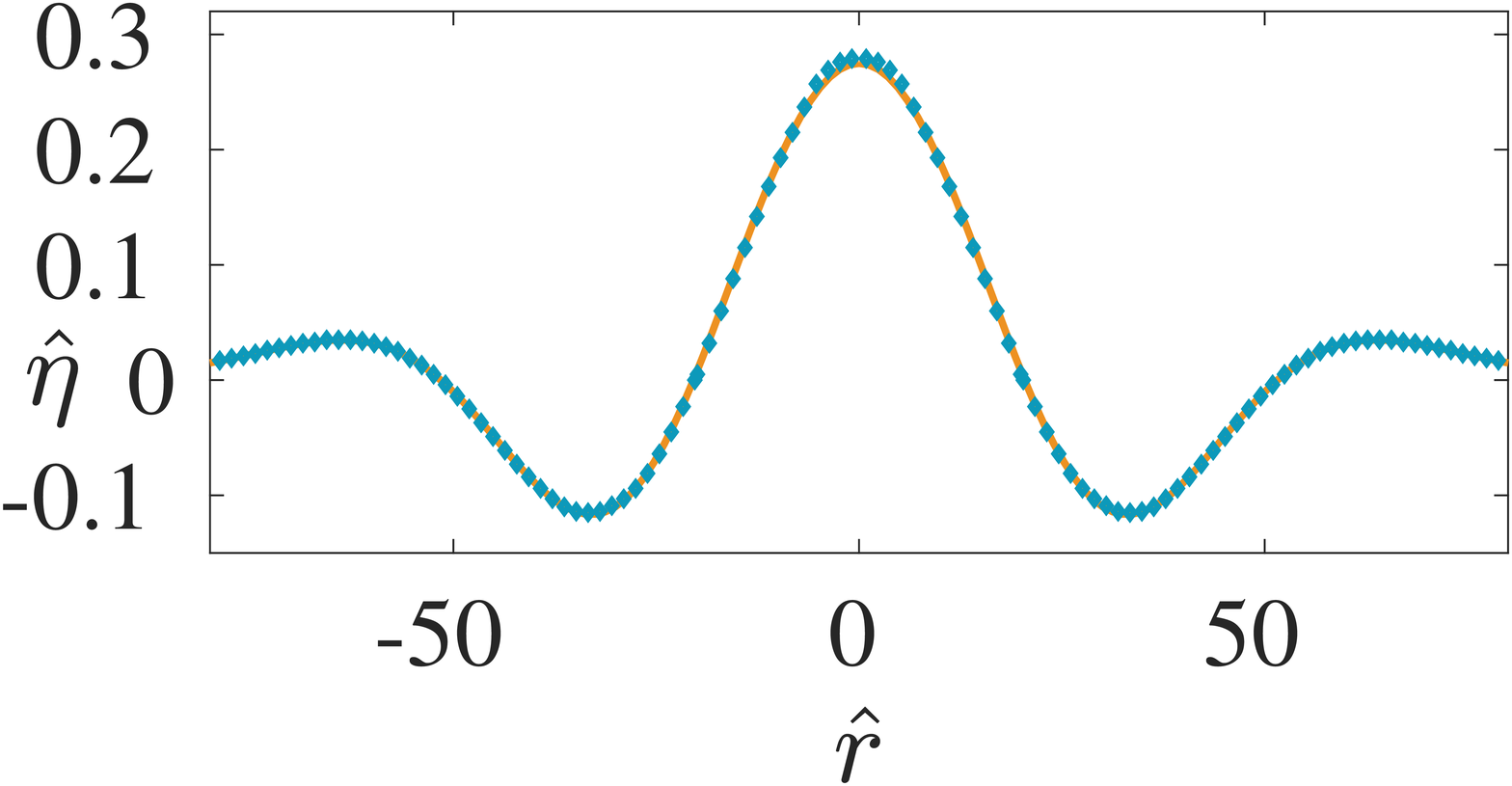}\label{fig_miles1_c}}\\	
	\caption{Evolution of a cavity from a hump of the intial form $\hat{\eta}(\hat{r},0)=\hat{a}_0\exp\left(-\frac{\hat{r}^2}{\hat{d}^2}\right)\left[1-\left(\frac{\hat{r}}{\hat{d}}\right)^2\right]$. In CGS units, $\hat{a}_0=0.54,\hat{d}=25, \beta \equiv \frac{\hat{a}_0}{\hat{d}}=0.0216 << 1$. Panel a) $\hat{t}=0.32$s when a cavity is formed, (inset) the initial perturbation. (Lin) Linear prediction obtained from expression \ref{1}. Panel b) Formation of a small bump at the symmetry axis at $\hat{t}=0.46$ s from the collapse of the cavity earlier at $\hat{t}=0.32$ s. Panel c) The bump at $\hat{r}=0$ does not rise to the same height as the initial perturbation (inset of panel (a)). In all three panels, note the excellent agreement between simulations and the linear prediction, \ref{1}. The physical parameters are chosen corresponding to air and water.}
	\label{fig_miles1}	
\end{figure}
\begin{figure}
	\centering
	\subfloat[$\hat{t}=0.320$ s]{\includegraphics[scale=0.14]{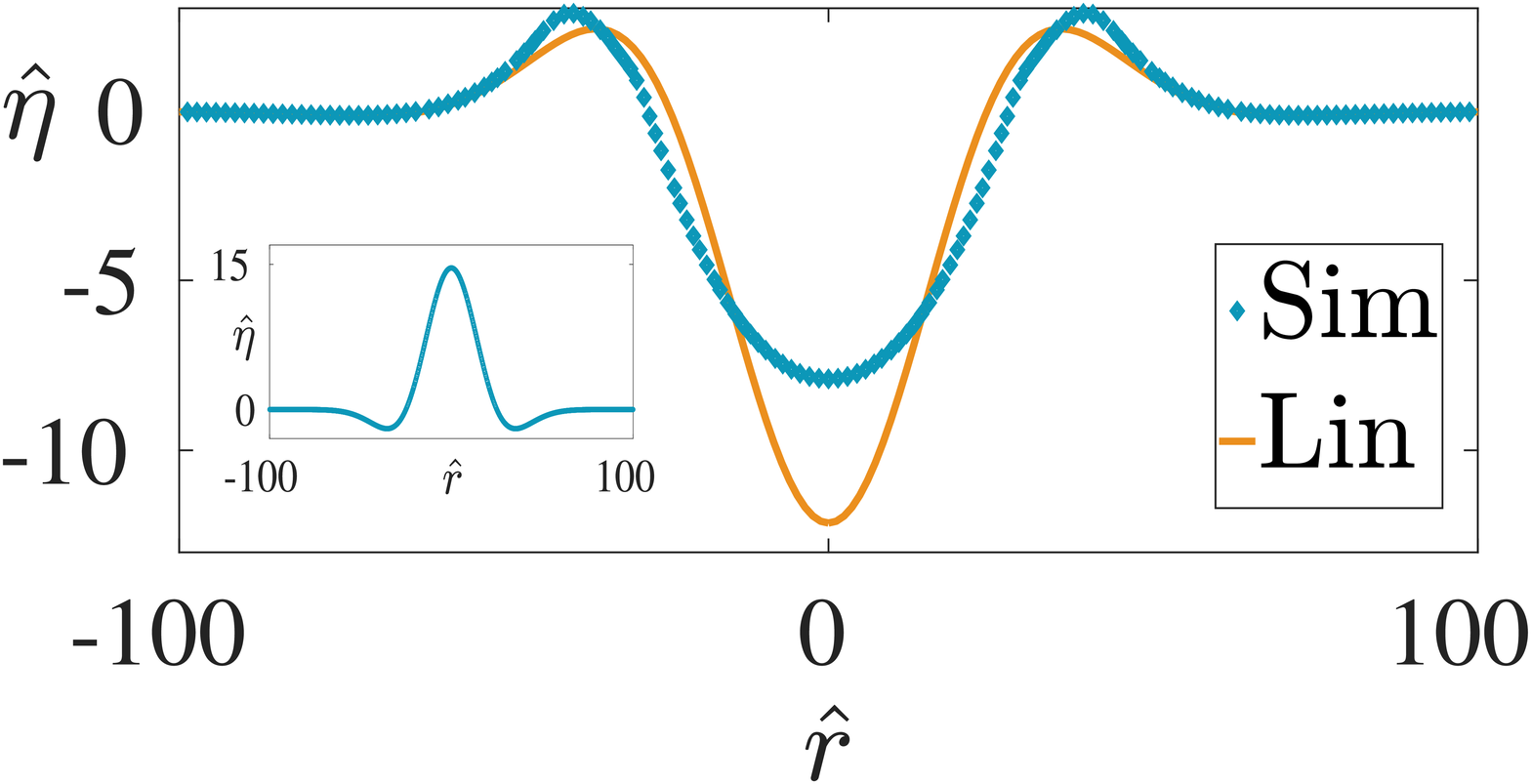}\label{fig_miles2_a}}
	\subfloat[$\hat{t}=0.5$ s]{\includegraphics[scale=0.14]{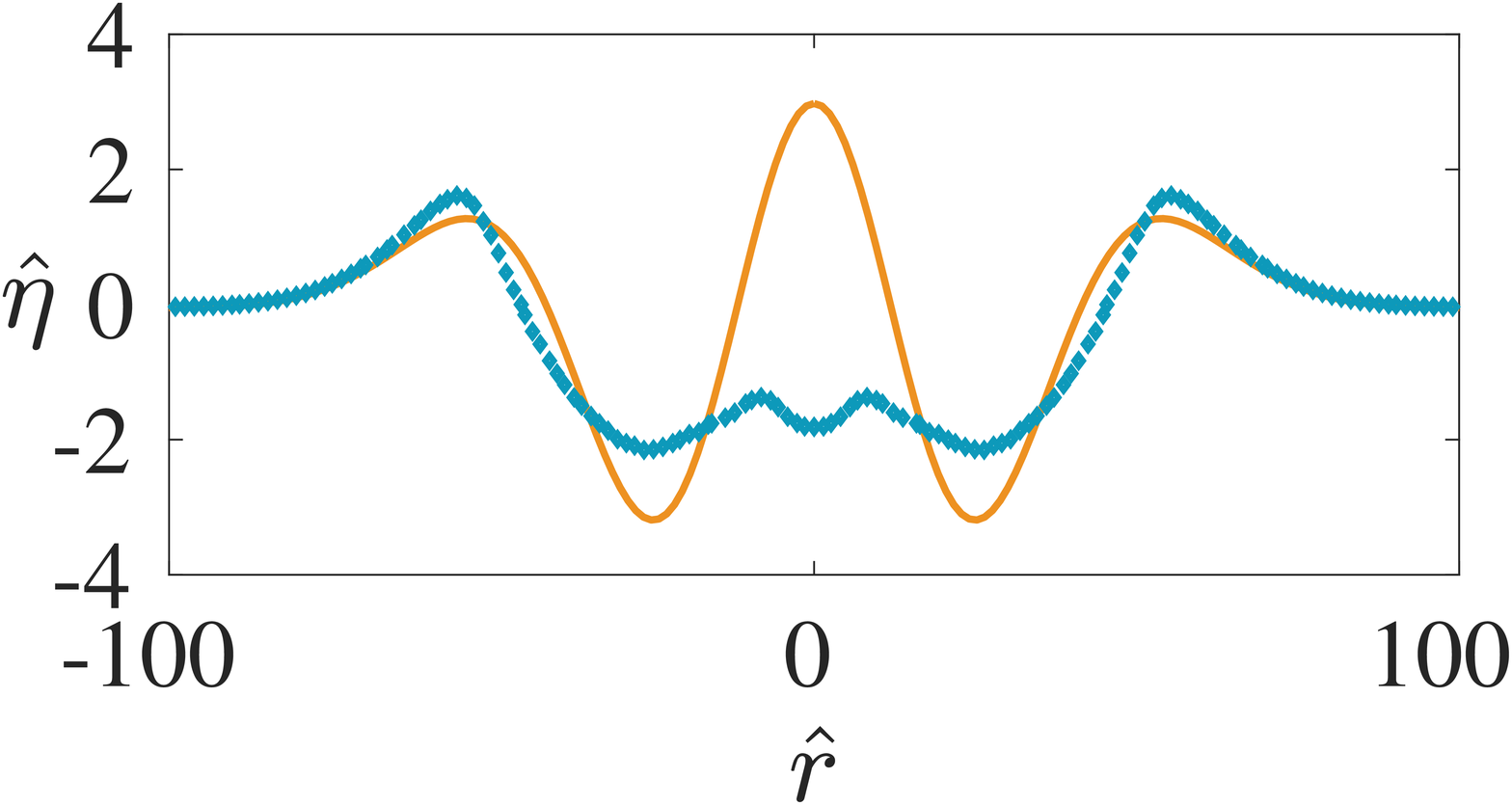}\label{fig_miles2_b}}\\
	\subfloat[$\hat{t}=0.73$ s]{\includegraphics[scale=0.14]{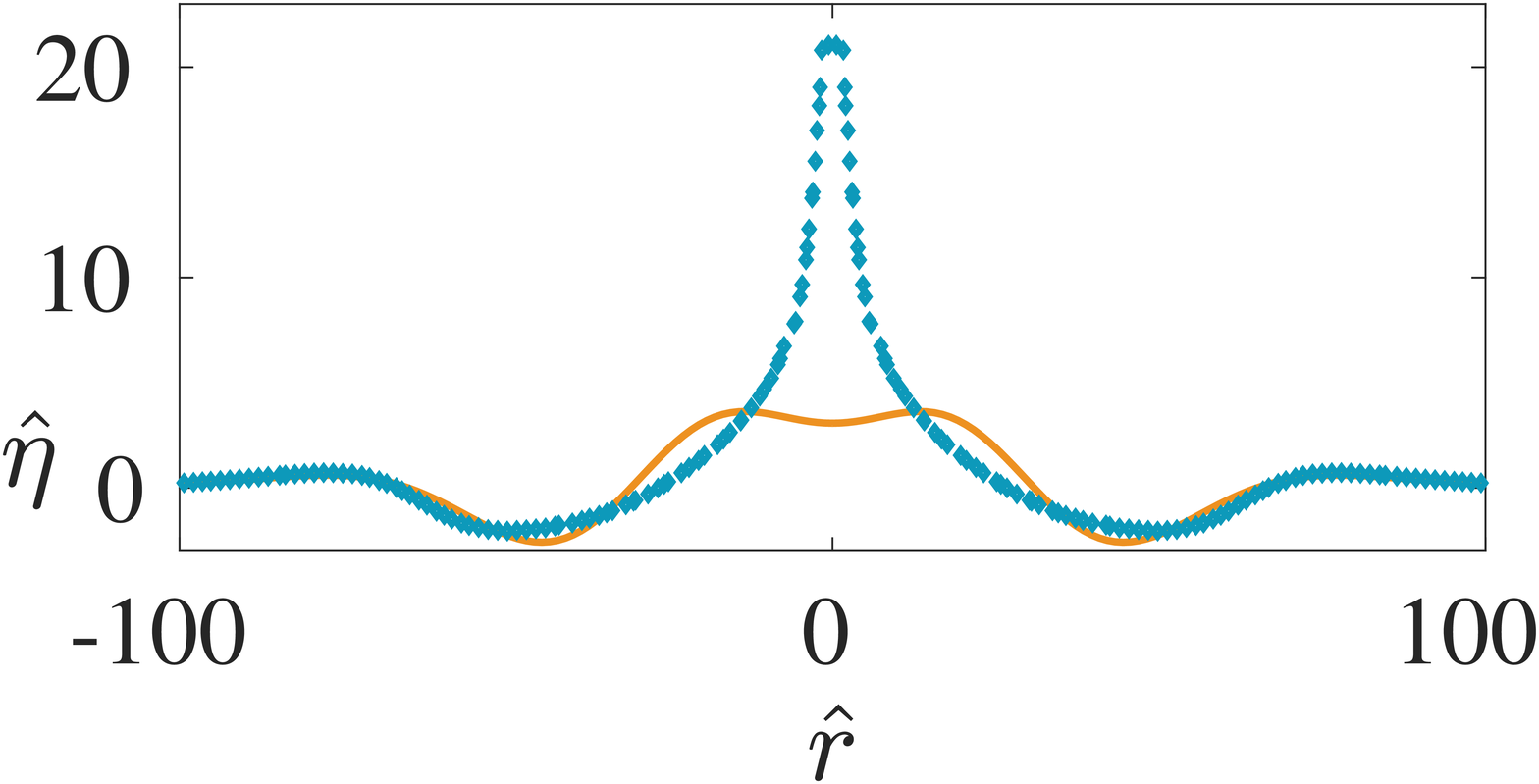}\label{fig_miles2_c}}\\
	\caption{Nonlinear evolution of a cavity starting from a hump of the same form  as fig. \ref{fig_miles1} with parameters (CGS) $\hat{a}_0=14.6, \hat{d}=25, \beta \equiv \frac{\hat{a}_0}{\hat{d}}=0.584 \sim \mathcal{O}(1)$. Panels (a), (b) and (c) depict the collapse of the cavity starting from a perturbation of much larger amplitude compared to fig. \ref{fig_miles1}. Note the formation of a slender jet in panel (c) which rises beyond the initial maximum amplitude $\hat{a}_0$.}
	\label{fig_miles2}	
\end{figure}

Fig. \ref{fig_cavity} depicts a cavity at an air-water interface (blue curve). This has been generated numerically by solving the incompressible Euler's equations with gravity (and zero surface tension) using Basilisk \citep{basilisk}, starting from an initial hump viz. $\hat{\eta}(\hat{r},0)=\hat{a}_0\exp\left(-\frac{\hat{r}^2}{\hat{d}^2}\right)\left[1-\left(\frac{\hat{r}}{\hat{d}}\right)^2\right],\; \hat{a}_0 > 0$ with characteristic height $\hat{a}_0$ and width $\hat{d}$ (see inset of fig. \ref{fig_miles1_a}). This initial condition was proposed by \cite{miles1968cauchy} and represents a volume conserving surface deformation. The linear dimensions of this perturbation (see fig. \ref{fig_cavity}) was chosen in the simulation to be far greater than the capillary-gravity length ($2.7$ mm) for air-water. The resultant cavity thus evolves dominantly under the influence of gravity with surface tension playing a relatively insignificant role, particularly in the early stage of collapse. When the non-dimensional ratio $\beta \equiv \dfrac{\hat{a}_0}{\hat{d}_0} << 1$, the collapse is a linear process governed by the CP solution. This is seen from the excellent agreement in figs. \ref{fig_miles1_a}, \ref{fig_miles1_b} and \ref{fig_miles1_c} between numerical simulations in Basilisk and linearised prediction from solving the integral in \ref{1} numerically. Note the formation of a bump like structure at the symmetry axis in fig. \ref{fig_miles1_b}, which eventually attains a maximum height less than $\hat{a}_0$, compare fig. \ref{fig_miles1_c} with the inset in fig. \ref{fig_miles1_a}. In contrast, when the value of $\beta$ is increased to $\mathcal{O}\left(1\right)$ the cavity collapse again generates a bump, but which now evolves into a thin, sharply shooting jet. This jet is shown in fig. \ref{fig_cavity} (red) as well as in more detail in figs. \ref{fig_miles2_a}, \ref{fig_miles2_b} and \ref{fig_miles2_c} where $\beta \approx 0.584$. Notably, this jet rises significantly higher ($\approx 20$ cms, see fig. \ref{fig_miles2_c}) compared to the initial hump height $\hat{a}_0=14.6$ cm. In this case, the temporal evolution is distinctly nonlinear as seen from the rather poor match seen in figs. \ref{fig_miles2_a}, \ref{fig_miles2_b} and \ref{fig_miles2_c} between the numerical simulations and the Cauchy-Poisson linear solution \ref{1} (indicated as `Lin' in figure legend). 

\subsection{Literature survey: Initial value problems (IVP)}
These numerical results presented in fig. \ref{fig_miles2}, emphasize the need for solving the nonlinear CP problem (i.e. predictions beyond eqn. \ref{1}) for a prescribed surface deformation $\hat{\eta}\left(\hat{r},0\right)$ and zero surface impulse $\hat{\phi}(\hat{r},\hat{z}=0,\hat{t}=0)=0$, particularly when one seeks a first principles understanding of the generation mechanism of the jet seen in fig. \ref{fig_miles2_c}. The theory literature, on solution to such IVPs particularly in axisymmetric coordinates is quite sparse with bulk of the rich analytical work, being focussed on understanding of \textit{time-periodic}, nonlinear, standing \citep{strutt1915deep} or travelling wave solutions \citep{stokes1847theory,stokes1880supplement}. Commencing from these seminal studies \citep{stokes1847theory,strutt1915deep}, a rich literature has since developed on finite amplitude, time-periodic surface waves, both in two dimensional coordinates \citep{penney1952part,taylor1953experimental,tadjbakhsh1960standing,fultz1962experimental,schwartz1981semi,schwartz1982strongly} as well as cylindrical, axisymmetric coordinates \citep{mack1962periodic,tsai1987numerical}. Additional, several studies of the stability of these finite amplitude solutions \citep{mercer1992standing,benjamin1967disintegration} have also been reported. 

Here our focus however, is not on these aforementioned time-periodic solutions but rather on those initial surface deformations which generate a sharply shooting jet, similar to the one seen in fig. \ref{fig_miles2_c}. An additional requirement is that the prescribed initial surface deformation $\hat{\eta}\left(\hat{r},0\right)$ should be simple enough to permit analysis in the nonlinear regime, at least perturbatively. The initial condition presented by \cite{miles1968cauchy} and discussed earlier in figs. \ref{fig_miles1} and \ref{fig_miles2} excites a continuum of modes (radially) initially. Extending the surface response due to such an initial perturbation beyond the CP linear regime described by \ref{1}, is technically demanding; this is mainly due to the necessity of accounting for interactions of a continuum of modes, interacting quadratically in a radial, axisymmetric geometry. In further analysis, we opt instead to study an alternative and apparently simpler initial condition in a radially confined geometry. The confinement assumption is mainly for analytical ease as it causes the radial part of the spectrum to be discrete instead of continuous.

\begin{figure}
	\centering
	\includegraphics[scale=0.27]{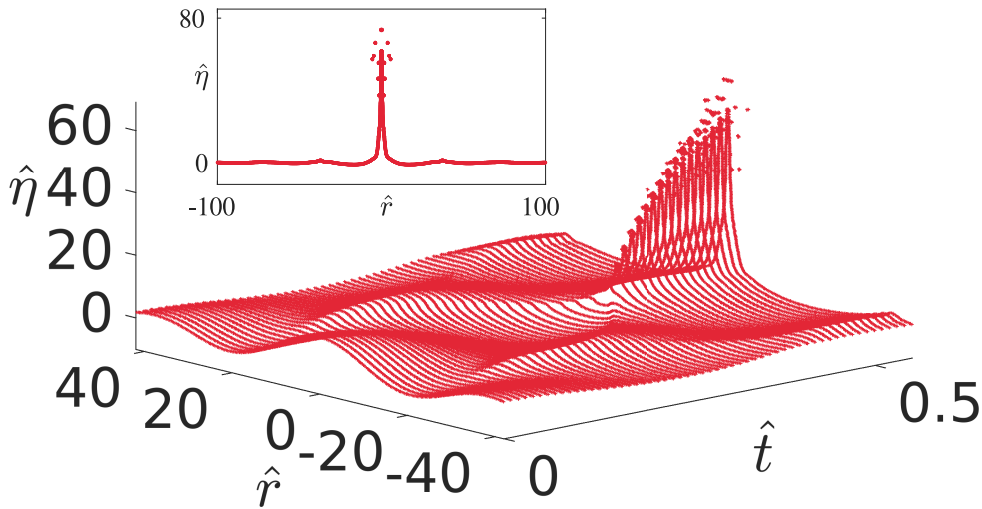}
	\caption{The development of a jet starting from a large scale surface deformation in the shape of single Bessel mode at $\hat{t}=0$ viz. $\hat{\eta}(\hat{r},0)=\hat{a}_0\mj_0\left(l_{35}\frac{\hat{r}}{\hat{R}_0}\right)$. The width of the initial hump is $\approx 40$ cms while the width of the radial domain is $\hat{R}_0=600$ cm, only a part of which is shown here. The resultant jet rises $> 60$ cms at its maximum, far exceeding the initial amplitude $\hat{a}_0=8.13$ cm. Note the formation of droplets from the tip of the jet close to $\hat{t}=0.5$ s. The jet profile at $\hat{t}=0.58$ s is shown in the inset. Parameters correspond to Case $5$ in table \ref{tab:kd}.}
	\label{fig_surface_eta}	
\end{figure}

In the initial condition that we study here, the free surface of a pool of very large depth, is deformed at time $\hat{t}=0$ as a single (radial) eigenmode to the cylindrical, axisymmetric, Laplacian operator viz. the zeroth order Bessel function. This surface deformation thus has the form $\hat{\eta}\left(\hat{r},0\right)=\hat{a}_0\mj_0\left(l_q\frac{\hat{r}}{\hat{R}_0}\right)$, $\hat{R}_0$ being the domain radius and $l_q$ ($q=1,2,\ldots$) a root of the Bessel function $\mathbb{J}_1\left(\cdot\right)$, this being necessary to satisfy no-penetration at the domain boundary. The length scales are chosen appropriately: the domain radius $\hat{R}_0$ as well as the width of the initial perturbation around the symmetry axis ($\approx \hat{R}_0l_q^{-1}$) are chosen to be quite large viz. $6$ metres and $\approx 0.5$ metres respectively (see fig. \ref{fig_surface_eta}). At sufficiently large $\hat{a}_0$, numerical simulation of the incompressible Euler's equation with gravity (and no surface tension) with this initial surface deformation, reveals the formation of a sharply shooting jet at the symmetry axis ($\hat{r}=0$) in fig. \ref{fig_surface_eta}. We notice the qualitative similarity of this jet to the one seen earlier in fig. \ref{fig_miles2}c viz. both rise far beyond their respective initial, maximum perturbation height $\hat{a}_0$. We will understand the mechanism of generation of this jet from first principles here by solving the corresponding Initial Value Problem (IVP).
In order to place our current study in perspective, we briefly summarise the literature on initial-value problems below. There are two classes of initial conditions for the solution to the linear CP problem viz. \\
\begin{itemize}
	\item  Waves are generated from fluid at rest and a specified \textit{initial surface deformation} (termed the primary CP problem recently by \cite{tyvand2021nonlinear}. This is of interest to us in this study. \\
	\item Waves and fluid motion are generated from a \textit{flat surface} due to an initial surface impulse (see expressions $3.2.10$ and $3.2.12$ in \cite{debnath1994nonlinear}). This is the secondary CP problem \citep{tyvand2021nonlinear}. We do not study this initial condition here.\\
\end{itemize}
As will be seen below, the secondary CP problem has received more attention compared to the primary one, possibly due to the experimental ease of starting with a flat surface rather than a deformed one. The primary and the secondary CP problem in two dimensional coordinates was investigated numerically using the potential flow equations by \cite{saffman1979note}. For the latter case, the authors applied a sinusoidal in space and time, pressure force for a certain duration on a flat interface, and reported the subsequent evolution of the free surface after this forcing was turned off. For the primary CP problem, they used the the finite amplitude solution of \cite{price1952finite} as their initial surface deformation. In both cases, they investigated waves of highest amplitude, obtaining good agreement with the well known experiments of G. I. Taylor \citep{taylor1953experimental}. In two-dimensional Cartesian coordinates, the primary CP problem was subsequently numerically investigated by \cite{longuet2001vertical_1} and \cite{longuet2001vertical_2}. The author(s) imposed a prescribed surface deformation in the shape of a circular trough, flanked by circular crests of larger radii. A sufficiently narrow trough as it collapsed in time, was numerically found to generate a jet rising upwards, see fig. $7$ in \cite{longuet2001vertical_2}. Notably, peak accelerations exceeding $10g$ were observed at the base of their cavity prior to the formation of the jet. In a followup work, the secondary CP problem was also investigated numerically by \cite{longuet2001breaking}. These authors numerically solved the IVP employing an initial condition, which corresponds to a standing wave solution to the linearised problem. The horizontal and vertical components of fluid motion in their simulation were prescribed in two dimensional coordinates, initially as $\hat{u}(\hat{x},\hat{z},0)=C\exp(k\hat{z})\cos(k\hat{x}), \hat{v}(\hat{x},\hat{z},0)=C\exp(k\hat{z})\sin(k\hat{x})$ with the surface being flat initially ($\hat{\eta}\left(\hat{x},0\right)=0$). For $C << 1$, a linear standing wave was observed. However at larger $C = O(1)$, the authors reported formation of a jet like structure at $x=0$ during the downward motion of the crest (see fig. $9b$ in \cite{longuet2001breaking}). Recently, the nonlinear, secondary CP problem in two dimensional Cartesian coordinates has also been solved analytically, employing small time expansion \citep{tyvand2021nonlinear,tyvand2021nonlinear_Lag}. These authors demonstrate significant differences between the linear and the nonlinear CP solution, with increasing amplitude of the initial pressure impulse. However a clear jet like structure analogous to what was presented in the numerical solution in \cite{longuet2001breaking}, is not discernable in their fig. $6$c \citep{tyvand2021nonlinear}. Of note, are also several experimental studies which have investigated the emergence of jets in setups which closely resemble the secondary CP problem (i.e. via application of surface impulse), mostly at capillarity dominated length scales. These include the studies by \cite{antkowiak2007short,bergmann2008origin,gordillo2020impulsive} of the tubular jet as well as several versions of the so-called Pokrowski's experiment \citep{lavrentiev1980effets}. 

The modal initial condition of interest to us in this study viz. $\hat{\eta}\left(\hat{r},0\right)=\hat{a}_0\mj_0\left(l_q\frac{\hat{r}}{\hat{R}_0}\right)$ corresponds to the primary CP problem described above \citep{tyvand2021nonlinear}, albeit in cylindrical, axisymmetric coordinates. This initial condition was first studied in \cite{farsoiya2017axisymmetric}, to analytically solve the viscous, linear, IVP at length scales (approx. a few centimetres) chosen such that surface tension as well as gravity were equally important. It was observed in the Direct Numerical Simulations (DNS) reported in \cite{farsoiya2017axisymmetric} that by systematically increasing the perturbation amplitude $\hat{a}_0$, a capillarity-gravity dominated jet emerged at the symmetry axis rising significantly higher than $\hat{a}_0$. The viscous, \textit{linear} theory presented in \cite{farsoiya2017axisymmetric} was unable to describe the formation of this jet or even account for its inception. In a subsequent study \citep{basak2021jetting}, the weakly non-linear solution to the IVP within an inviscid framework and accurate upto $\mathcal{O}\left(\epsilon^2\right)$ $\left(\epsilon \equiv \frac{\hat{a}_0 l_q}{\hat{R}_0}\right)$ corresponding to the same initial condition as \cite{farsoiya2017axisymmetric} was developed. Here too the length scales of interest were similar to that of \cite{farsoiya2017axisymmetric} with gravity and surface tension forces being equally strong and both forces were accounted for, in the theory. This nonlinear theory \citep{basak2021jetting} was a significant improvement over the linear model of \cite{farsoiya2017axisymmetric} and was able to predict the inception of the jet, comparing well with inviscid simulations of the Euler's equations with gravity and surface tension \citep{basak2021jetting}. The physical mechanism underlying this jet formation was however not presented in \cite{basak2021jetting}. Additionally, for given choice of $l_q$, surface tension and gravity, the analytical theory in \cite{basak2021jetting} becomes singular for certain values of the domain size $\hat{R}_0$, this being related to triadic internal resonance. It was demonstrated \citep{basak2021jetting} that these singularities owed their origin to the presence of both surface tension as well as gravity in the theoretical model and represented the cylindrical counterparts of such singularities, better known in the context of Wilton ripples \citep{wilton1915lxxii} in two dimensional coordinates. 

Now, it is already known from experiments on bubble bursting at millimetric \citep{tagawa2012highly,deike2018dynamics,gordillo2019capillary} and micron length scales \citep{lee2011size}, that jets accompany such cavity collapse quite routinely at these scales where gravity is insignificant compared to surface tension during the collapse. Motivated partly by this observation, simulations of the incompressible, Euler's equation with only surface tension (no gravity) with the aforementioned initial condition of \cite{farsoiya2017axisymmetric} have been reported recently in \cite{kayal2022dimples}. It was demonstrated clearly in these simulations, that surface tension alone at sufficiently small scales can produce a jet similar to what was observed at much larger capillary-gravity length scales in \cite{farsoiya2017axisymmetric} and \cite{basak2021jetting}. Concomitantly, the weakly non-linear solution to the IVP for this initial condition and taking into account only surface tension has also been developed in \cite{kayal2022dimples}. Comparison of this analytical theory against numerical simulations demonstrated very good agreement \citep{kayal2022dimples}, additionally also shedding light into the physical mechanism at work driven by gradients of curvature. It was proven \citep{kayal2022dimples} that in order to describe the inception of the jet in the surface tension driven case, one needs to account for nonlinear effects of curvature (the gradient of which drives the flow). To capture this accurately, it was shown that the analytical theory needs to be at least third order accurate i.e. $\mathcal{O}(\epsilon^3)$. This third order accurate solution was reported in \cite{kayal2022dimples} and has been found to be free from the aforementioned internal resonance related singularities in \cite{basak2021jetting}, thereby also alleviating a practical deficiency of the capillary-gravity model studied earlier in \cite{basak2021jetting}. 

Our present study treats the converse case of \cite{kayal2022dimples} considering large amplitude surface waves under the infuence of gravity (and no surface tension). To be consistent we choose our length scales of interest to be typically in the metre range (radial domain size is $6$ m) and this represents a scale up of nearly two orders of magnitude in length, as compared to all our earlier studies \citep{farsoiya2017axisymmetric,basak2021jetting,kayal2022dimples}. The motivation to search for jets at such large scales, comes from the rogue wave literature where a very interesting recent experimental observation of \cite{mcallister2022wave} report a `spike wave'. \cite{mcallister2022wave} generate a spike wave which rises up to $6$ m in height, via focussing of waves at the symmetry axis of a cylindrical pool. Note that in their case, the focussing was achieved not via cavity collapse (which generically also tends to focus waves at the symmetry axis) but via wavemakers at the periphery of a very large cylindrical tank ($25$ m dia) which generated wave components arranged to be in phase at the symmetry axis, leading to spike wave there. Given our numerical observation of purely capillarity driven jets (sans gravity) for the surface deformation $\hat{\eta}(\hat{r},\hat{t}=0)=\hat{a}_0\mj_0(l_q\frac{\hat{r}}{\hat{R}_0})$ at millimetric length scales in \cite{kayal2022dimples}, we ask here if gravity acting alone, at typical length scales of tens of metres may also produce a jet similar to the one seen in \cite{kayal2022dimples}? We will demonstrate here, theoretically and computationally, that the answer to this question is in the affirmative and also explain the physical mechanism at work. Analogous to the purely surface tension driven, nonlinear theory developed in \cite{kayal2022dimples}, the purely gravity driven, second-order, non-linear theory developed here is devoid of singularities, arising from internal resonance. This theory developed via multiple scale analysis is then compared to numerical simulations obtaining very good agreement.
\section{Physical mechanism of jet formation: wave focussing}
\begin{figure}
	\centering
	\subfloat[]{\includegraphics[scale=0.28]{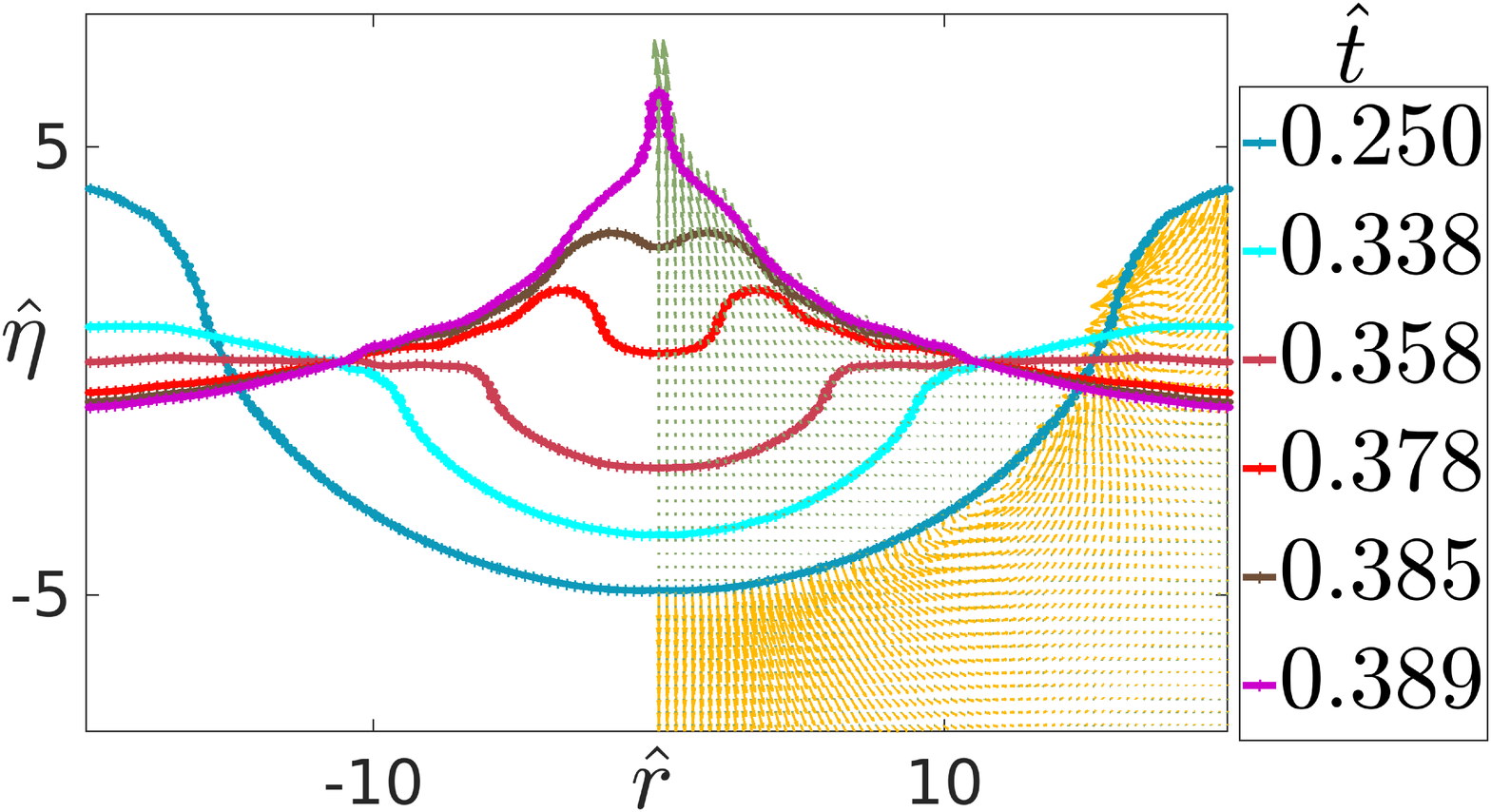}}\\
	\subfloat[$\hat{t}=0.285$ ]{\includegraphics[scale=0.13]{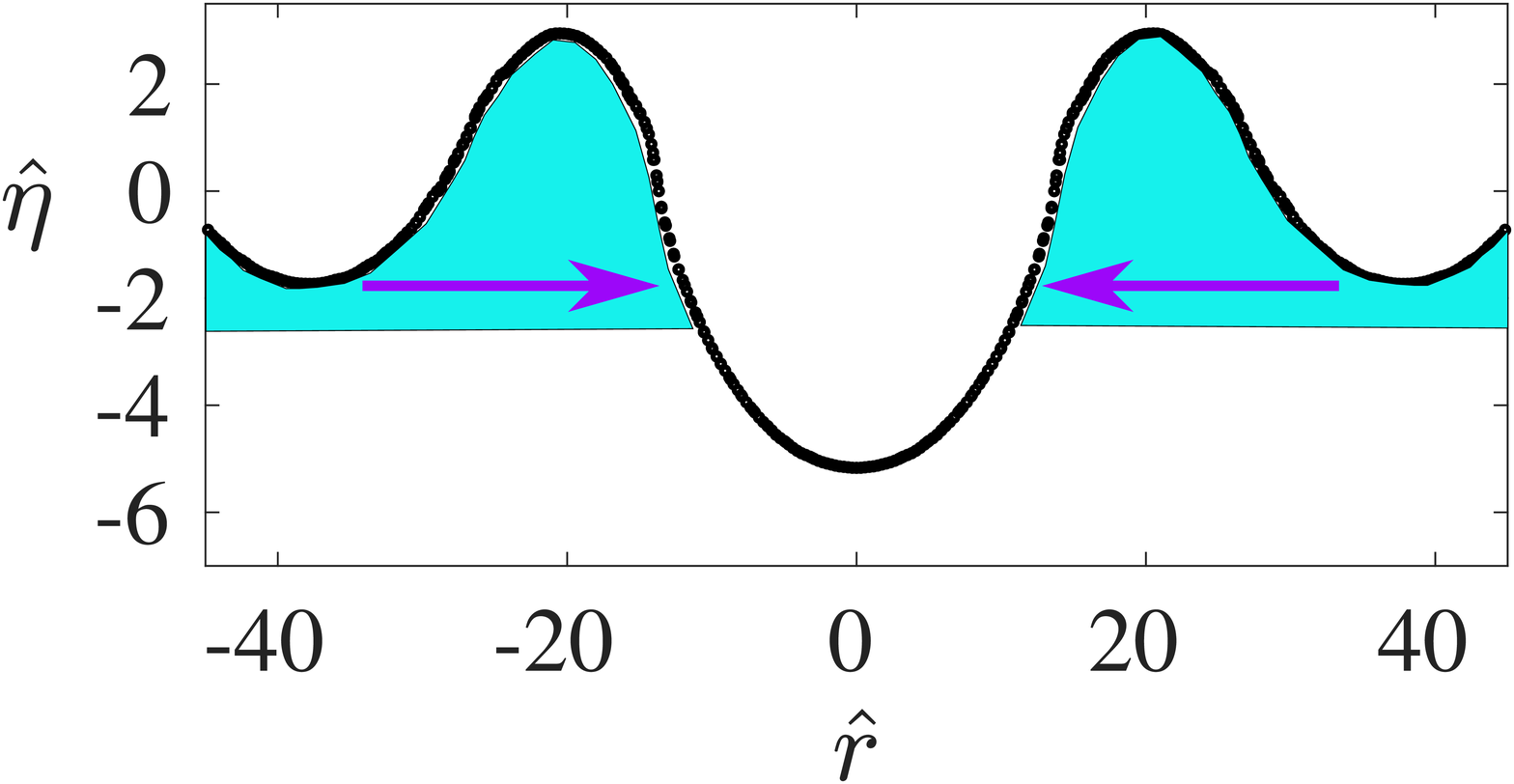}\label{fig_focussing_a}}
	\subfloat[$\hat{t}=0.35$ s]{\includegraphics[scale=0.13]{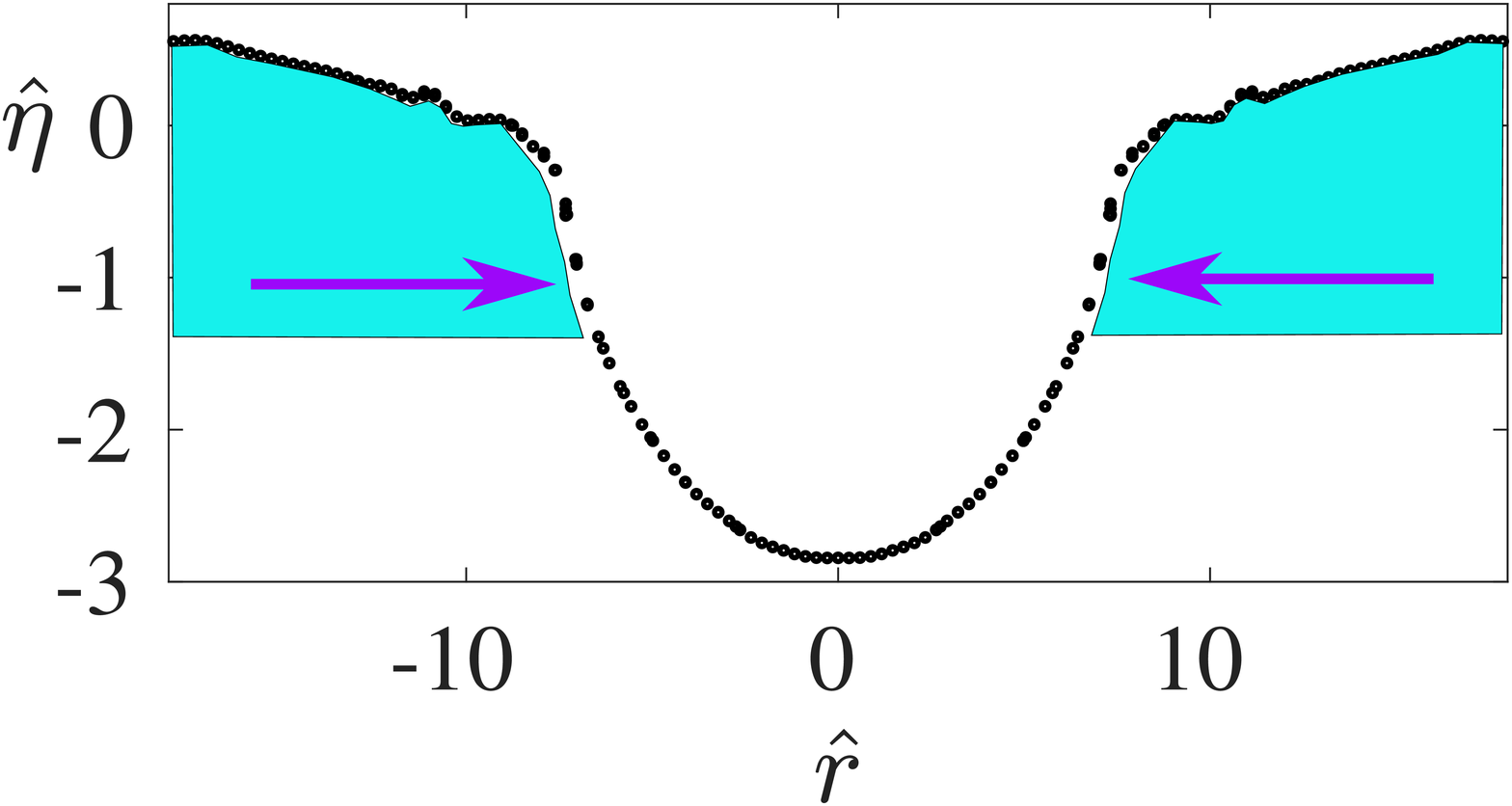}\label{fig_focussing_b}}\\
	\subfloat[$\hat{t}=0.38$ s]{\includegraphics[scale=0.13]{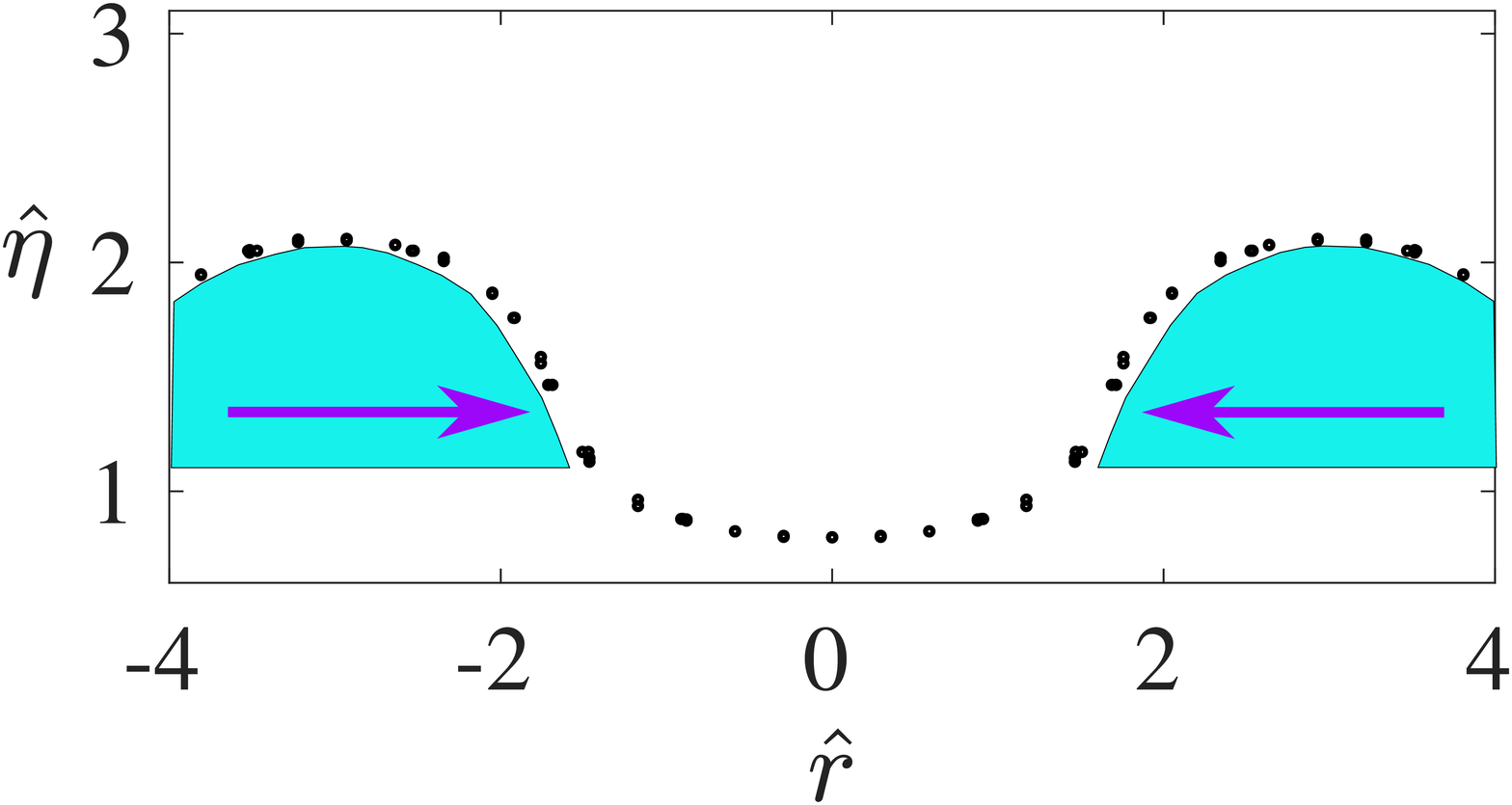}\label{fig_focussing_c}}
	\subfloat[$\hat{t}=0.39$ s]{\includegraphics[scale=0.13]{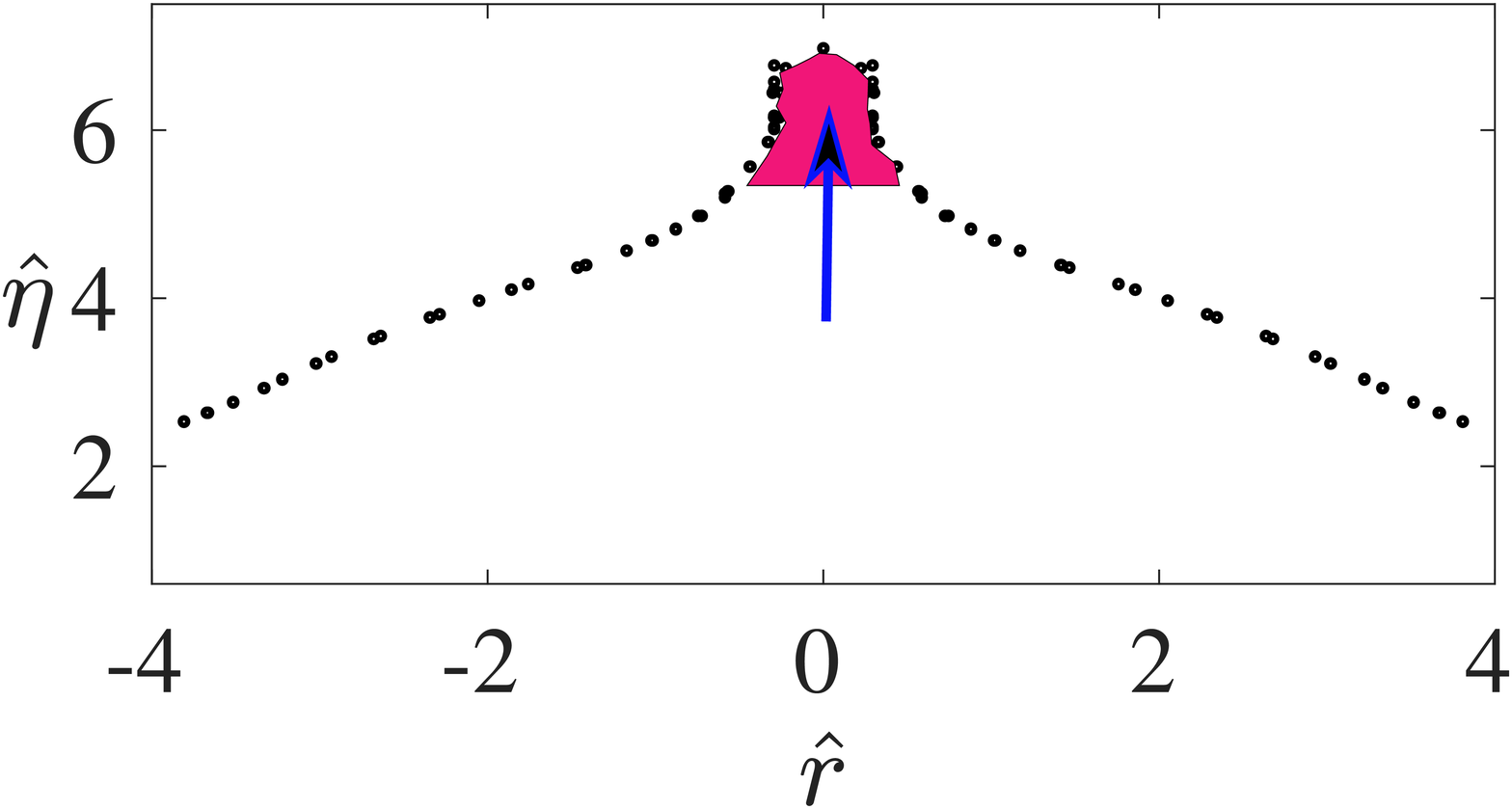}\label{fig_focussing_d}}
	\caption{Panel (a) depicts the temporal evolution and development of a jet at the symmetry axis. This is from simulations with $\epsilon=1.5$ (Case $5$ in table \ref{tab:kd}). The arrows (yellow) indicate the velocity field at $\hat{t}=0.25$ s while (green) arrows are at $\hat{t}=0.389$ s. Panels (b)-(d) show the same simulation as panel (a), providing a closer view of the inward focussing of the concentric, large, amplitude surface-gravity wave towards the symmetry axis. The inward motion of the crests (in blue) leads to the eventual emergence of the jet at $\hat{t}\approx 0.389$ s.}
	\label{fig_focussing}
\end{figure}

It is useful to visualize the cavity collapse and the subsequent jet that arises with the modal initial condition $\hat{\eta}\left(\hat{r},0\right)=\hat{a}_0\mj_0\left(l_q\frac{\hat{r}}{\hat{R}_0}\right)$ when $\hat{a}_0>0$ is sufficiently large. Referring to the jet in Fig. \ref{fig_surface_eta} , we note that this rises to a height exceeding $0.6$ m, seven times higher than the initial perturbation amplitude, $\hat{a}_0 \approx 8.13$ cm. The cavity shown in fig. \ref{fig_focussing}, panel (a) in deep blue is from the same simulation as fig. \ref{fig_surface_eta}. The initial perturbation is chosen to have $\epsilon=1.5 > 1$  implying that this is a highly nonlinear jet. The radial inward arrows in fig. \ref{fig_focussing}, panels (b)-(d) indicate the radial inward motion of the wave crests (shaded in blue) as time progresses. Following the figures one notes the emergence of the jet at $\hat{t}\approx 0.39$ s.

\begin{figure}
	\centering
	\subfloat[$\hat{t}=0.01$ s, $\epsilon=0.1$ ]{\includegraphics[scale=0.23]{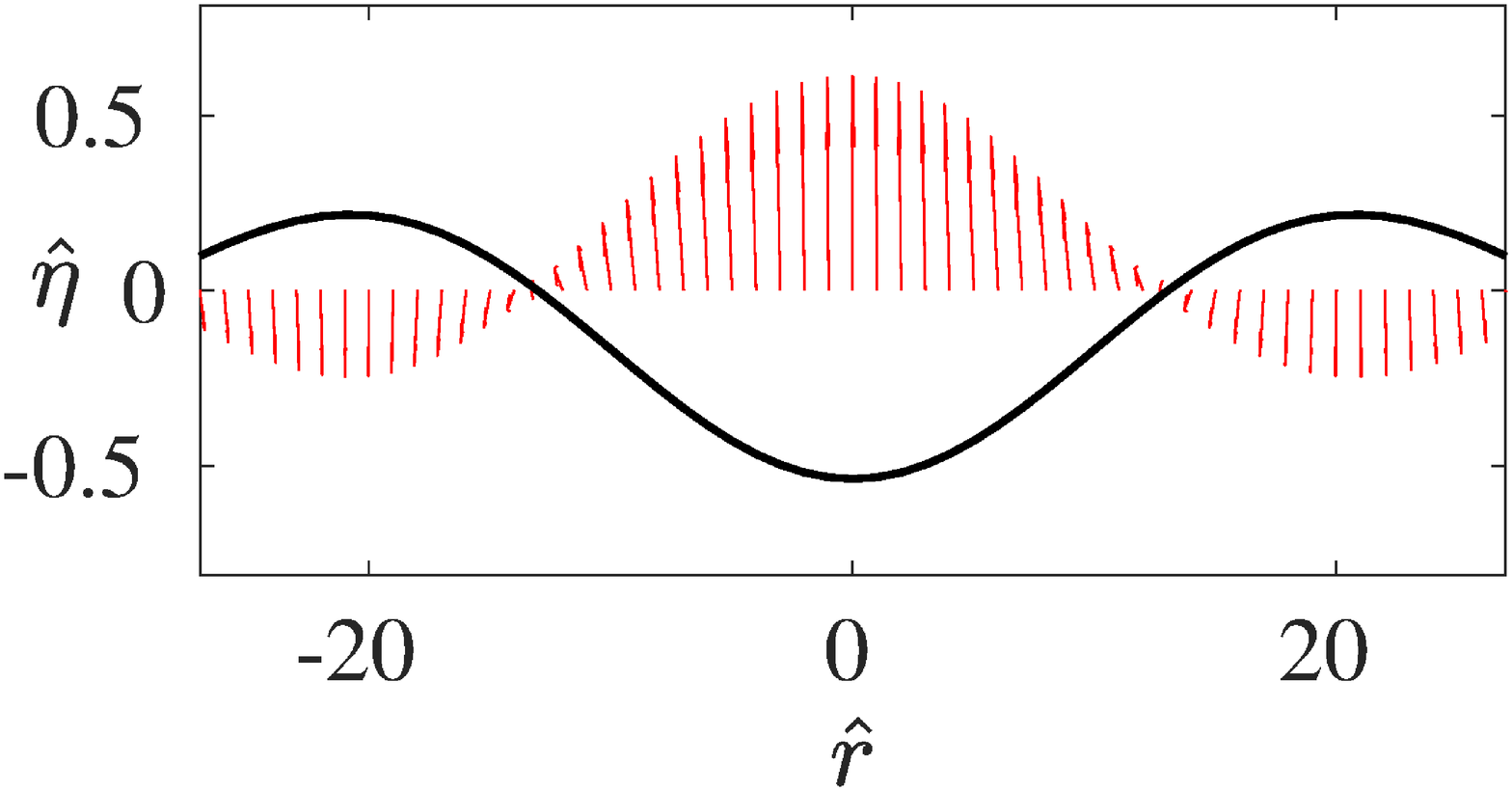}\label{fig_press_vel_a}}\\
	\subfloat[$\hat{t}=0.285$ s, $\epsilon \equiv 0.1$]{\includegraphics[scale=0.2]{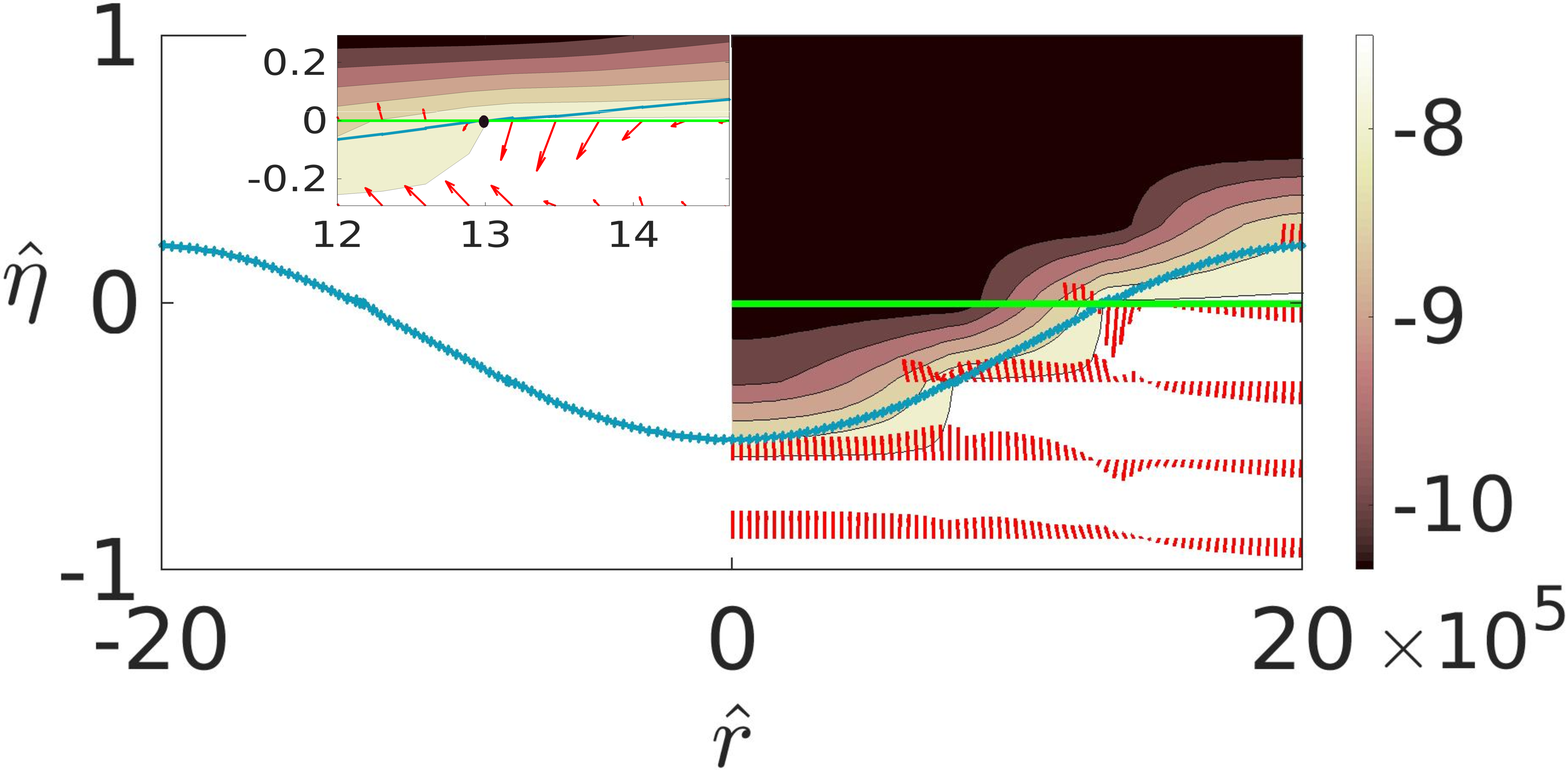}\label{fig_press_vel_b}}\\
	\subfloat[$\hat{t}=0.245$ s, $\epsilon \equiv 1.5$]{\includegraphics[scale=0.2]{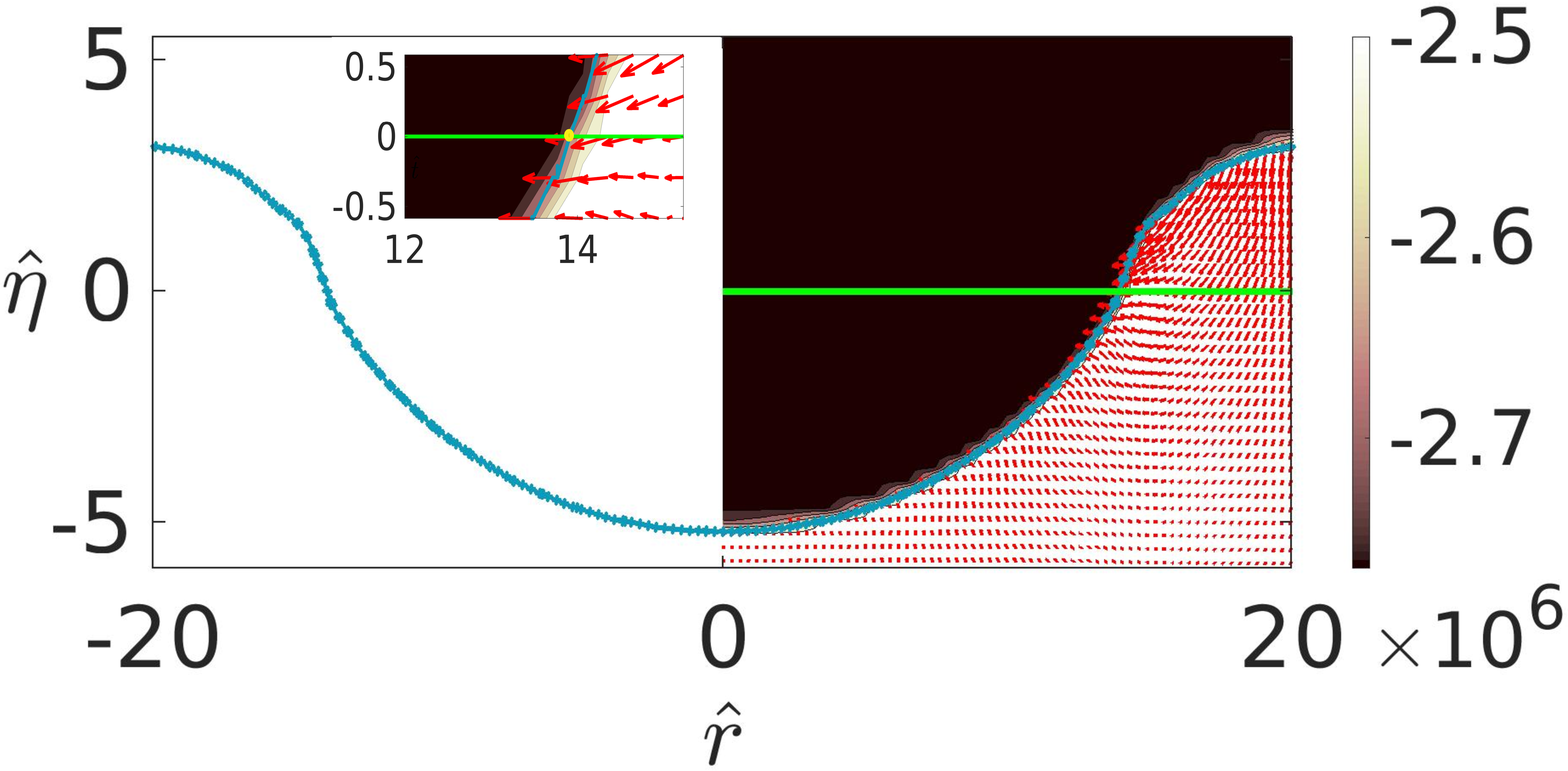}\label{fig_press_vel_c}}
	\caption{Panel (a): Arrows represent the total velocity vector at $\hat{z}=0$ predicted from eqns. \ref{eq2}b and \ref{eq2}c, the interface (black curve) is obtained from eqn. \ref{eq2}a at $\hat{t}=0.01$. Panels (b) and (c): Numerical simulation of a cavity developed from the initial perturbation $\hat{\eta}=\hat{a}_0\mj_{0}\left(l_q\dfrac{\hat{a}_0}{\hat{R}_0}\right)$. Panel (b) shows a cavity (deep blue curve) developing from a small amplitude initial perturbation satisfying $\hat{a}_0 << \hat{R}_0l_q^{-1}$. This cavity generates a standing wave as predicted by linear theory in equations \ref{eq2}a, with (near) zero velocity at $\hat{r} = \hat{r}^{*}$. The location of the first zero crossing of the perturbed interface, is indicated as black dot in the inset. Panel (c) represents a steep walled cavity (deep blue curve) arising from a perturbation of the same form as panel (b) but of large amplitude satisfying $\hat{a}_0 > \hat{R}_0l_q^{-1}$. Note that in contrast to panel (b), the zero crossing of the interface ($\hat{r} = \hat{r}^{*}$ indicated as an yellow dot in the inset) has a strong, inward horizontal velocity component. The background contours in panels (b) and (c) are for pressure. The fluorescent green line in both panels (b) and (c) represent the undisturbed water level while arrows indicate the instantaneous velocity field.}
	\label{fig_press_vel}	
\end{figure}

In order to understand physically the radially inward motion seen in figs. \ref{fig_focussing}a we shall employ mass conservation via the kinematic boundary condition in the description below. Recall that the solution to the linear CP problem provided in expression \ref{1} for the modal surface deformation viz. $\hat{\eta}(\hat{r},\hat{t}=0)=-\hat{a}_0\mj_0\left(l_q\frac{\hat{r}}{\hat{R}_0}\right)$ and $\hat{\phi}\left(\hat{r},\hat{z}=0,0\right)=0$ ($\hat{a}_0 > 0$) is 
\begin{subequations}\label{eq2}
	\begin{align}        
		&\hat{\eta}(\hat{r},\hat{t}) = -\hat{a}_0\mj_0\left(l_q\frac{\hat{r}}{\hat{R}_0}\right)\cos\left(\sqrt{l_q \frac{g}{\hat{R}_0}}\hat{t}\right),\tag{\theequation a}\\
		& \hat{u}_z(\hat{r},\hat{z},\hat{t}) =\frac{\hat{a}_0l_q^{1/2}g^{1/2}}{R_0^{1/2}}\mj_0\left(l_q\frac{\hat{r}}{\hat{R}_0}\right)\exp\left(l_q\frac{\hat{z}}{\hat{R}_0}\right)\sin\left(\sqrt{l_q \frac{g}{\hat{R}_0}}\hat{t}\right),\tag{\theequation b}\\
		&\hat{u}_r(\hat{r},\hat{z},\hat{t}) =-\frac{\hat{a}_0l_q^{1/2}g^{1/2}}{R_0^{1/2}}\mj_1\left(l_q\frac{\hat{r}}{\hat{R}_0}\right)\exp\left(l_q\frac{\hat{z}}{\hat{R}_0}\right)\sin\left(\sqrt{l_q \frac{g}{\hat{R}_0}}\hat{t}\right), \tag{\theequation c}
	\end{align}         	
\end{subequations}
where $\hat{u}_r$ and $\hat{u}_z$ are the radial and axial components of perturbation velocity. Here a \textit{negative sign} is deliberately used in the prescription for $\hat{\eta}(\hat{r},\hat{t}=0)$ to generate an initial trough (instead of a hump) around $\hat{r}=0$. The linearised kinematic boundary condition viz.  $\left(\dfrac{\partial\hat{\eta}}{\partial\hat{t}}\right) = \hat{u}_z\left(\hat{z}=0,\hat{r},\hat{t}\right)$, implies that the temporal evolution of the interface (at linear order) is determined only by the vertical component of perturbation velocity $\hat{u}_z$ at the undisturbed liquid level $\hat{z}=0$. Thus for this initial condition, the zeroes of the initial surface deformation viz. the solution to $\mj_0$$\left(l_q\frac{\hat{r}}{\hat{R}_0}\right)=0$ behave as nodes (at linear order). Let us denote the first zero of $\mj_0$$\left(l_q\frac{\hat{r}}{\hat{R}_0}\right)=0$ as $\hat{r}^{*}$. The linearised kinematic condition predicts that at $\hat{r}=\hat{r}^{*}$,
\begin{eqnarray}
	\left(\frac{\partial \hat{\eta}}{\partial \hat{t}}\right)_{\hat{r}=\hat{r}^{*}} = 0,\; \text{at all time}. \label{eq2.2}
\end{eqnarray}
Crucially, note from eqns. \ref{eq2}b and \ref{eq2}c that although the value of $\hat{u}_z(\hat{r} = \hat{r}^{*},\hat{z}=0,\hat{t})= 0$, the value of $\hat{u}_r(\hat{r} = \hat{r}^{*},\hat{z}=0,\hat{t} < \frac{T}{4})\neq 0$, being a small negative value in the first quarter of the oscillation cycle with time period $T$, viz. the early stages of cavity collapse. Thus linear theory predicts a small \textbf{\textit{radially inward}}, horizontal velocity at $\hat{r}=\hat{r}^{*}$, as seen from the negative sign of \ref{eq2}c for $\hat{r}=\hat{r}^{*},\hat{z}=0$ and $\hat{t} < \frac{T}{4}$. However this radially inward velocity, cannot affect the time evolution of the interface at linear order as the (linearised) kinematic boundary condition depends only on the vertical perturbation velocity $\hat{u}_z$ at $\hat{z}=0$. Consequently, linear theory predicts that for this modal initial condition and with $\hat{a}_0$ sufficiently small, a zero of $\mj_{0}(\cdot)$ will behave as a \textit{node} at all time (eqn. \ref{eq2.2}). A standing wave with nodes predicted from linear theory thus develops and no radial movement of the nodes is expected. This is depicted in fig. \ref{fig_press_vel_a} where the arrows indicate the velocity vector at $\hat{z}=0$ with magnitude $\sqrt{\hat{u}_r^2 + \hat{u}_z^2}$ (from eqns. \ref{eq2}b and \ref{eq2}c) evaluated at $\Delta{\hat{t}=0.01} s$. It is seen that the vertical velocity is zero at the first zero crossing of the interface, although there is a small negative horizontal component of velocity at $\hat{r}=\hat{r}^{*}$ (not visible at this scale). Fig. \ref{fig_press_vel_b} provides support to the aforementioned argument from numerical simulations of the Euler's equation for small $\hat{a}_0$. Note that the cavity seen in this figure has been generated starting from a hump viz. $\hat{\eta}(\hat{r},\hat{t}=0)=\hat{a}_0\mj_0\left(l_q\frac{\hat{r}}{\hat{R}_0}\right), \hat{a}_0 > 0$, although in a linear description this is not important. As this is a small amplitude simulation ($\hat{a}_0 << \hat{R}_0l_q^{-1}$), the initial perturbation generates a linear standing wave. The inset depicts the instantaneous zero of $\mj_{0}(\cdot)$ as a black dot, and it is seen that the velocity is nearly zero at this location, implying no inward movement of the two crests seen in the figure in the small amplitude limit. 

As $\hat{a}_0$ is increased, this behaviour of the interface around the zero crossing $\hat{r}=\hat{r}^{*}$ changes qualitatively, with nonlinearity becoming important. The exact nonlinear kinematic boundary condition at $\hat{r}=\hat{r}^{*}$ may be written as,  
\begin{eqnarray}
\left(\dfrac{\partial\hat{\eta}}{\partial\hat{t}}\right)_{\hat{r}=\hat{r}^{*}} = \left[\hat{u}_z\left(\hat{z}=\eta,\hat{r},\hat{t}\right) - \hat{u}_r\left(\hat{z}=\eta,\hat{r},\hat{t}\right)\left(\frac{\partial\hat{\eta}}{\partial \hat{r}}\right)\right]_{\hat{r}=\hat{r}^{*}}. \label{eq2.3}
\end{eqnarray}
Now as a rough approximation, let us evaluate the right hand side of \ref{eq2.3} using the linearised expressions \ref{eq2.2} at $\hat{z}=0$ (instead of $\hat{z}=\hat{\eta}$). If we start with a trough initial condition as shown in fig. \ref{fig_press_vel_a}, in the first quarter of the oscillation cycle ($\hat{t} <  T/4$), at the zero crossing $\hat{r}=\hat{r}^{*}$ the interface slope $\left(\frac{\partial\hat{\eta}}{\partial \hat{r}}\right)_{\left(\hat{r}=\hat{r}^{*},\hat{t} <  T/4\right)} >0$. As $\hat{u}_z(\hat{z}=0, \hat{r}=\hat{r}^{*},\hat{t} < T/4) =0$ and  $\hat{u}_r(\hat{z}=0, \hat{r}=\hat{r}^{*},\hat{t} < T/4) < 0$ from eqn. \ref{eq2}c, eqn. \ref{eq2.3} thus predicts a \textit{positive upward displacement} of the interface at the nodal point viz. $\left(\dfrac{\partial\hat{\eta}}{\partial\hat{t}}\right)_{\left(\hat{r}=\hat{r}^{*}, \hat{t} < T/4\right)} > 0$. It is easy to see that an upward displacement of the interface at $\hat{r}=\hat{r}^{*}$, implies that the zero crossing location moves inwards towards the axis of symmetry. Note that this prediction crucially depends on taking into account the second non-linear term on the right hand side of eqn. \ref{eq2.3} without which there is no such inward motion. It will be seen in the next section that a more precise, weakly nonlinear theory predicts this inward motion of the interface zero crossing, in quantitative agreement with numerical simulations. 

So far, our justification for the flow focussing has been purely kinematic (using only mass conservation) without appealing to forces or pressure gradients, which drive the flow. Incidentally, this provides insight into why one should generically expect jets from such large amplitude modal perturbations, independent of whether one is investigating the phenomena at tiny length scales (surface tension driven as in \cite{kayal2022dimples}) or far larger, gravity driven scales as is our present case. In order to invoke dynamics, we once again refer to fig. \ref{fig_press_vel}b and \ref{fig_press_vel}c which contrast the linear behaviour against the nonlinear one respectively. As we do not consider surface tension or the upper fluid in our current model (in simulations, the interface separates air above and water below which are well separated in density), the interface may be treated as a free-surface along which pressure remains constant at all time. Thus the pressure-gradient vector $\bm{\nabla}\hat{p}$ points normal to the interface indicated in deep blue in both these figures. Consider the direction of the pressure gradient vector at $\hat{r}=\hat{r}^{*}$ in fig. \ref{fig_press_vel}c (graphically the point $\hat{r}=\hat{r}^{*}$ is the intersection of the blue curve  with the undisturbed liquid level line indicated in fluorescent green, the point being indicated as a black and yellow dot respectively, in the inset in figures \ref{fig_press_vel}b and \ref{fig_press_vel}c respectively). It is clear that the vector $\bm{\nabla}\hat{p}$ in the nonlinear case of fig. \ref{fig_press_vel}c is aligned nearly horizontally at $\hat{r}=\hat{r}^{*}$, pointing radially outwards. The resultant acceleration and flow then, is driven nearly radially inwards and towards the axis of symmetry leading to the inward motion of the two humps shown earlier in fig. \ref{fig_focussing}. The vertical jet at the axis of symmetry is then understood to be a result of flow focussing due to mass conservation. We also point out the contrast of  background pressure contour levels between figs.  \ref{fig_press_vel}b  and \ref{fig_press_vel}c. These contours clearly indicate that the magnitude of the pressure gradient is significantly larger in the nonlinear case, producing in the latter case nearly horizontal velocity vectors around the zero crossing which causes inward motion of the crests. 

The flow focussing that we explain above from first principles, can cause such jets to appear due to cavity collapse. It is clear from the above discussion that this is a generic mechanism operating not only in axi-symmetric geometry where it is expectedly more intense due to the $\frac{1}{r}$ divergence, but also in two dimensional Cartesian geometry where an initial prescribed large amplitude surface deformation, may have an axis of symmetry towards which such focussing may be directed. The physical mechanism related to such flow focussing was remarked upon by \cite{longuet2001breaking}, see page $496$ end of first paragraph. 
While the numerical study by \cite{longuet2001breaking} investigated the planar counterparts of these jets, our current study represents the cylindrical axisymmetric, purely gravity driven counterpart which has not been studied analytically or computationally before. We also mention related studies by \cite{longuet1995critical,longuet1997critical} where the authors model the potential flow due to a sink of strength $S$ (to model flow convergence due to wave focussing) moving upwards with velocity $V$ (to model the jet). The free surface (i.e. surface of constant pressure) may be chosen to be Dirichlet ellipsoids or hyperboloids, and the jets generated in their model are particularly relevant to the formation of such jets from spherical bubbles. 

\section{Weakly nonlinear theory: multiple scale approach}
In this section, we develop an $\mathcal{O}(\epsilon^2)$ weakly nonlinear theory employing the method of multiple scales and $\epsilon \equiv \frac{\hat{a}_0l_q}{\hat{R}_0}$ as perturbation parameter, in order to predict from first principles how does the jet develop. Note that in our earlier studies \citep{basak2021jetting,kayal2022dimples}, we have used the closely related method of strained coordinates (Lindstedt-Poincare technique). Our usage of multiple-scales here is motivated by the fact that at sufficiently early time, the effect of nonlinearity should be small and the interface should behave as a linear standing wave. At a longer time window however, the effect of nonlinearity becomes apparent including the production of free and bound mode components absent initially. We derive here the amplitude equations governing the modulation of the primary mode as this was not obtained earlier in the strained coordinate method \citep{basak2021jetting,kayal2022dimples}. As our study is restricted to jets on a water pool only, for pure water the corresponding viscous-gravity length-scale $\left(\nu^2/g\right)^{1/3}<< 1$ mm is quite small compared to the length scales of interest to us here. Consequently, we resort to the inviscid, irrotational, potential flow approximation as is standard practise in analysing large-scale surface gravity waves \citep{mcallister2022wave}. These potential flow equations are,
\begin{eqnarray}
\left(r,z, \eta\right) \equiv \frac{l_q}{\hat{R}_0}\left(\hat{r},\hat{z},\hat{\eta}\right), \quad t \equiv {\left(l_q\frac{g}{\hat{R}_0}\right)}^{1/2}\hat t, \quad\phi \equiv {\left(\frac{ l_q^3}{\hat{R}_0^3g}\right)}^{1/2}\hat \phi
\end{eqnarray}
which yields  the following set of equations written in axisymmetric, cylindrical coordinates. Here the (non-dimensional) $\phi$ represents the disturbance velocity potential and (non-dimensional) $\eta$ represents the disturbed free surface. The governing equations and boundary conditions are:
\begin{subequations}\label{eq3.2}
	\begin{align} 
&\nabla^2\phi = \frac{\partial^2\phi}{\partial r^2} + \frac{1}{r}\frac{\partial\phi}{\partial r} + \frac{\partial^2\phi}{\partial z^2} = 0,\tag{\theequation a}\\
&\eta + \left(\frac{\partial \phi}{\partial t} + \frac{1}{2} \left(\frac{\partial \phi}{\partial r}\right)^2 + \frac{1}{2}\left(\frac{\partial \phi}{\partial z}\right)^2\right)_{z=\eta}= 0,\quad\quad \frac{\partial\eta}{\partial t} + \left(\frac{\partial \eta}{\partial r}\frac{\partial \phi}{\partial r} - {\frac{\partial \phi}{\partial z}}\right)_{z=\eta}=0,\tag{\theequation b,c}\\
&\left(\phi\right)_{z\rightarrow -\infty} \rightarrow \text{finite},\quad \left(\frac{\partial \phi}{\partial r}\right)_{r = l_q} = \left(\frac{\partial \eta}{\partial r}\right)_{r = l_q} = 0,\tag{\theequation d,e,f}\\
& \int_{0}^{l_q} dr \hspace{1mm} r\eta(r,t) = 0,\tag{\theequation g}\\
& \text{with initial conditions}\quad\eta(r,0) = \epsilon \mathrm{J_0}(r), \quad \frac{\partial \eta}{\partial t}(r,0) = 0, \quad \phi(r,z,0) = 0. \tag{\theequation h,i,j}
\end{align} 
\end{subequations}
Eqn. \ref{eq3.2} a is the Laplace equation in cylindrical axisymmetric coordinates, \ref{eq3.2} b,c are the constant pressure condition (from Bernoulli equatoin) and the kinematic boundary condition respectively, both at the free surface. Eqn. \ref{eq3.2}d is the finiteness of the velocity potential at infinite depth (deep water limit is assumed for simplcity). \ref{eq3.2}e,f are the no-penetration and free-edge boundary conditions at the outer radial boundary $\hat{r}=\hat{R}_0$. Eqn. \ref{eq3.2}g is the overall mass conservation while \ref{eq3.2}h,i and j are initial conditions corresponding the the primary CP problem. 

The number $l_q$ ($q \in \mathbb{Z}^{+}$) in the initial condition $\hat{\eta}(\hat{r},\hat{t}=0)=\hat{a}_0\mj_0\left(l_q\frac{\hat{r}}{\hat{R}_0}\right)$ satisfies $\mj_1(l_q)=0$. The chosen integer value of $q$ in $\epsilon \equiv l_q\frac{\hat{a}_0}{\hat{R}_0}$, is related to the number of extremas of $\mj_{0}$ within the radial domain $\hat{R}_0$. The numerical value of $\hat{R}_0l_q^{-1}$ provides a rough measure of the wavelength of the initial perturbation while non-linearity is controlled by the magnitude of the perturbation amplitude $\hat{a}_0$ relative to $\hat{R}_0l_q^{-1}$. In order to minimise the effect of the confining radial boundary $\hat{R}_0$ on the cavity collapse process, we require $\hat{R}_0l_q^{-1} << \hat{R}_0$ and this is ensured by choosing $l_q >> 1$. In this study, we choose $q=35$ implying $l_{35}=110.74$. 
The variables $\phi$, $\eta$ and $t$ are expanded in a power series in $\epsilon$ ($\epsilon \ll 1$) for finite $l_q$. Employing mutiple scale analysis, we replace the temporal dependencies in all dependent variables as $T_n \equiv \epsilon^n t$ (upto second order). Thus we have upto second order,
\begin{eqnarray}
&&\phi(r,z,T_0,T_2) = \epsilon\phi_1(r,z,T_0,T_2) + \epsilon^2\phi_2(r,z,T_0,T_2)  + \mathcal{O}(\epsilon^3)\label{eq3.3}\\
&&\eta(r,t) = \epsilon\eta_1(r,T_0,T_2) + \epsilon^2\eta_2(r,T_0,T_2)  + \mathcal{O}(\epsilon^3) \label{eq3.4}
\end{eqnarray}
where $T_0\equiv t$ and $T_2\equiv\epsilon^2t$. After performing a Taylor series expansion of equations \ref{eq3.2}b and \ref{eq3.2}c about the unperturbed interface $z = 0$ and thereafter substituting expansions \ref{eq3.3} and \ref{eq3.4} in equations \ref{eq3.2}a-j, we extract the following set of linear equations governing $\phi_i(r,z,T_0,T_2)$ and $\eta_i(r,T_0,T_2)$ at the $i^{th}$ order for all $i = 1,2,3,\ldots$. Hence at every $\mathcal{O}(\epsilon^i)$ and using the short-hand symbol $D_n \equiv \frac{\partial}{\partial T_n}$, we have
\begin{subequations}\label{eq3.5}
\begin{align} 
&\nabla^2\phi_i=0 \tag{\theequation a}\\
&D_0\phi_i +\eta_i = \mathscr{N}_i(r,T_0,T_2), \quad  \frac{\partial \phi_i}{\partial z} - D_0\eta_i = \mathscr{M}_i(r,T_0,T_2) \quad \text{at}\;z = 0 \tag{\theequation b,c}\\
&\left(\phi_i\right)_{z\rightarrow -\infty} \rightarrow \text{finite},\quad\left(\frac{\partial \phi_i}{\partial r}\right)_{r = l_q} = 0,\quad \left(\frac{\partial \eta_i}{\partial r}\right)_{r = l_q} = 0, \tag{\theequation d,e,f}\\
&\int_{0}^{l_q} dr \hspace{1mm} r\eta_i(r,T_0,T_2) = 0 \tag{\theequation g}\\
\textrm{and}\quad& \eta_i(r,0,0) = \delta_{1i}\mathrm{J_0}(r), \quad D_0\eta_i(r,0,0) = 0, \quad \phi_i(r,z,0,0) = 0 \tag{\theequation h,i,j}
\end{align} 
\end{subequations}
Here $\delta_{1i}$ is the Kronecker delta while $\mathscr{M}_i(r,T_0,T_2)$ and $\mathscr{N}_i(r,T_0,T_2)$ are  nonlinear terms involving products of lower order solutions of $\phi_i$ and $\eta_i$ with $\mathscr{M}_1(r,T_0,T_2)=\mathscr{N}_1(r,T_0,T_2)=0$. Due to this at the linear order we have a homogeneous set of equations. The expressions for $\mathscr{M}_2(r,T_0,T_2),\mathscr{N}_2(r,T_0,T_2)$ as well as $\mathscr{M}_3(r,T_0,T_2),\mathscr{N}_3(r,T_0,T_2)$ (necessary for eliminating resonant forcing) are provided in the appendix. As all $\phi_i$ satisfy the Laplace equation \ref{eq3.5}a, we expand $\phi_i$ and $\eta_i$ at every order in a \textit{Dini series} \citep{mack1962periodic} in the following form (note that $q$ is a given fixed integer used to indicate the primary Bessel mode that is excited initially) 
\begin{eqnarray}
&&\phi_i(r,z,T_0,T_2) = \sum_{j = 0}^{\infty}p_i^{(j)}(T_0,T_2)\exp(\alpha_{j,q} z)\mathrm{J_0}(\alpha_{j,q} r) \label{dini_phi} \label{eq3.6}\\
\textrm{and}&&\eta_i(r,z,T_0,T_2) = \sum_{j = 0}^{\infty}a_i^{(j)}(T_0,T_2)\mathrm{J_0}(\alpha_{j,q} r), \quad \alpha_{j,q} \equiv \frac{l_j}{l_q}, \; i=1,2,3,\; j=1,2,3,4\ldots\dots \nonumber\\\label{eq3.7}
\end{eqnarray}
By construction, each term in the expansion in \ref{eq3.6} satisfies the Laplace equation \ref{eq3.5}a while each term in \ref{eq3.7} satisfies the mass conservation equation \ref{eq3.5}g. In addition, \ref{eq3.6} and \ref{eq3.7} together respect the finiteness condition as well as the no-penetration and the free-edge conditions \ref{eq3.5}d,e,f. Our task thus reduces to ensuring that eqns. \ref{eq3.5}b,c are satisfied, and these in turn determine equations governing $p_{i}^{(j)}(T_0,T_2)$ and $a_{i}^{(j)}(T_0,T_2)$ in the expansions \ref{eq3.6} and \ref{eq3.7}. These equations have to be solved subject to initial conditions \ref{eq3.5}h,i,j.

Substituting \eqref{eq3.6} and \eqref{eq3.7} into \ref{eq3.5}b,c, taking inner products with $\mj_{0}(\alpha_{n, q}) r dr$ and using the orthogonality relations for Bessel functions (see supplementary material in \cite{basak2021jetting} for the relevant orthogonality relations), we obtain the following equations at each $\mathcal{O}(\epsilon^i)$:
\begin{subequations}\label{eq3.8}
	\begin{align} 
&\left(D_0^2 + \omega_{j,q}^2 \right)p_i^{(j)}(T_0,T_2) = \frac{2}{l_q^2 \mathrm{J_0}^2(l_j)}\int_{0}^{l_q} dr\, r \hspace{1mm} \mathrm{J_0}(\alpha_{j,q} r) \left\{D_0\mathscr{N}_i(r,T_0,T_2) + \mathscr{M}_i(r,T_0,T_2) \right\}, \tag{\theequation a}\\
& \text{and } a_i^{(j)}(T_0,T_2) = \frac{2}{l_q^2 \mathrm{J_0}^2(l_j)}\int_{0}^{l_q} dr \hspace{1mm} r\,\mathrm{J_0}(\alpha_{j,q} r) \mathscr{N}_i(r,T_0,T_2) - D_0p_i^{(j)}(T_0,T_2) \tag{\theequation b}
\end{align} 
\end{subequations}
Here $\omega_{j,q}^{2} \equiv \alpha_{j,q}$ is the non-dimensional form of the dispersion law for pure gravity waves on a free-surface. Eqns. \ref{eq3.8}a is a second order, partial differential equation which is solved subject to the following initial conditions (these are obtained by substituting \ref{eq3.6} and \ref{eq3.7} into initial conditions \ref{eq3.5}h,i,j)
\begin{eqnarray}\label{eq3.9}
&& p_i^{(j)}(0,0) = 0, \quad 
D_0p_i^{(j)}(0,0) = \frac{2}{l_q^2 \mathrm{J_0}^2(l_j)}\int_{0}^{l_q} dr \hspace{1mm}r\, \mathrm{J_0}(\alpha_{j,q} r) \mathscr{N}_i(r,0,0) - \delta_{1i}\delta_{qj} \nonumber \\
\end{eqnarray}
The expression for $a_i^{(j)}\left(T_0,T_2\right)$ may then be obtained from equation \ref{eq3.8}b, knowing the expression for $p_i^{(j)}\left(T_0,T_2\right)$ from the solution to \ref{eq3.8}a. 
\subsection{Linear Solution ($i=1$)}
At $\mathcal{O}(\epsilon)$, $\mathscr{N}_i(r,T_0,T_2)=\mathscr{M}_i(r,T_0,T_2)=0$ which reduces eqn. \eqref{eq3.8}a to a homogeneous equation along with initial conditions obtained from \ref{eq3.9} for $i=1$ viz.
\begin{eqnarray}
p_1^{(j)}(0,0) = 0, \quad  D_0p_1^{(j)}(0,0) = -\delta_{qj}\label{eq3.10}
\end{eqnarray} 
This leads to 
\begin{eqnarray}
&&p_1^{(q)}(T_0,T_2) = \left(\mu_{1q}(T_2)\cos T_0+\nu_{1q}(T_2)\sin T_0\right)\label{eq3.11}\\
\textrm{and}&& a_1^{(q)}(T_0,T_2) = \left(\mu_{1q}(T_2)\sin T_0-\nu_{1q}(T_2)\cos T_0\right)\label{eq3.12} \\
&& p_1^{(j)}(T_0,T_2) = 0, \quad a_1^{(j)}(T_0,T_2) = 0\quad  \forall j\neq q \nonumber
\end{eqnarray}
$\mu_{1q}(T_2)\text{ and }\nu_{1q}(T_2)$ in \ref{eq3.11} and \ref{eq3.12} are functions of the slow time-scale $T_2$ and satisfy initial conditions $\mu_{1q}(0)=0,\,\nu_{1q}(0)=-1$, obtained from \eqref{eq3.10}. Hence at $\mathcal{O}(\epsilon)$, $\phi_1\,\text{ and }\eta_1$ are obtained as
\begin{eqnarray}
&&\phi_1(r,z,T_0,T_2) = \left(\mu_{1q}(T_2)\cos T_0+\nu_{1q}(T_2)\sin T_0\right)\mathrm{J_0}(r)\exp (z)\label{eq3.13}\\
\textrm{and}&&\eta_1(r,T_0,T_2) = \left(\mu_{1q}(T_2)\sin T_0-\nu_{1q}(T_2)\cos T_0\right)\mathrm{J_0}(r)\label{eq3.14}
\end{eqnarray}
Note that if we were to terminate our calculation at this order, then $\mu_{1q}$ and $\nu_{1q}$ would be treated as constants being $\mu_{1q}=0,\nu_{1q}=-1$ and substituting these in \ref{eq3.13} and \ref{eq3.14}, we obtain the standing wave that is predicted at linear order viz.
\begin{eqnarray}
	\phi(r,z,T_0) = -\epsilon\sin(T_0)\mj_{0}(r)\exp(z),\quad \eta(r,z,T_0) = \epsilon\cos(T_0)\mj_{0}(r) \label{eq3.15}
\end{eqnarray}
Eqns. \ref{eq3.15} predict that the interface $\eta$ evolves temporally as a standing wave with unit frequency (in non-dimensional sense). Expectedly, at all time there is only the primary mode and no transfer of energy to modes other than this mode (index $q$) is predicted at this order. Consequently, there is no jet seen in the time evolution of the standing wave. At the axis of symmetry, the initial perturbation represented by $\eta$ in eqn. \ref{eq3.15} has an amplitude $\epsilon$ at $T_0=0$ and after one time period at $T_0=2\pi$, once again rises to the same amplitude at this location. While we have seen earlier that the radial velocity at the node ($r = r^{*}$) is non-zero and radially inward already at this order, this velocity does not affect the time evolution of the standing wave. In order to theoretically predict, the radial inward movement of nodes which leads to the jet seen in fig. \ref{fig_surface_eta}, our calculation must be extended to nonlinear order. We expect to obtain amplitude equations governing $\mu_{1q}(T_2)$ and $\nu_{1q}(T_2)$ appearing in eqns. \ref{eq3.13} and \ref{eq3.14} by going upto nonlinear order and this is presented next.
\subsection{Nonlinear Solution ($i=2,3$)}
On a longer time-scale $T_2$, the amplitude of the primary mode ($j=q$) gets modulated, this being dictated by nonlinear equations governing $\mu_{1q}(T_2)$ and $\nu_{1q}(T_2)$. In order to determine these equations, it becomes necessary to obtain expressions upto $\mathcal{O}(\epsilon^3)$ for $\mathscr{M}_2$ and $\mathscr{N}_2$ as well as $\mathscr{M}_3$ and $\mathscr{N}_3$), where resonant forcing of the primary mode is encountered and eliminated. The nonlinear calculation proceeds in an analogous manner to the linear one outlined above, although the algebra, expectedly, becomes increasingly tedious. We outline only the important steps here. 

For $i=2$, $\mathscr{M}_2(r,T_0,T_2)\neq 0$ and $\mathscr{N}_2(r,T_0,T_2)\neq 0$ and expressions for these are provided in the appendix. As a result, equation \ref{eq3.8}a governing $p_2^{(i)}(T_0,T_2)$ is an inhomogenous equation. To prevent very lengthy calculation for solving this equation, we introduce an approximation as follows. While solving eqn. \ref{eq3.8}a for $p_2^{(j)}(T_2)\;\left(j=0,1,2,3,\ldots\right)$, constants of integration appear which strictly speaking, are not constants but functions of the slow time scale $T_2$. These govern the modulation of the amplitude of the secondary modes. In order to determine the equations governing this modulation, one needs to go to higher order (presumably beyond third order). Our purpose here is to obtain a nonlinear approximation which contains enough physics to model the inception of the jet seen earlier. Thus in order to prevent very lengthy, higher order calculations we ignore the temporal variation of these functions treating them instead as constants to be determined from initial conditions. Our analytical solution thus contains the modulation of the amplitude of the linear solution (i.e. the primary mode with index $q$), but neglects amplitude modulation of the secondary modes generated in the spectrum, via nonlinearity. The justification for this will be obtained \textit{a posteriori}, when we compare our analytical predictions with numerical simulations. After lengthy algebra, the $\mathcal{O}(\epsilon^2)$ corrections are found to be,  
\begin{eqnarray}
&&\phi_2(r,z,T_0,T_2) = \left[\frac{\nu_{1q}(T_2)^2-\mu_{1q}(T_2)^2}{2}\sin \left(2T_0\right)+\mu_{1q}(T_2)\nu_{1q}(T_2)\cos \left(2T_0\right)\right]\mathrm{J_0}^2(l_q )\nonumber\\
&& +\sum_{n = 1}^{\infty}\left[\xi_{n,q}^{(2)}\sin(\omega_{n,q}T_0) + \xi_{n,q}^{(3)}\left(\mu_{1q}(T_2),\nu_{1q}(T_2)\right)\cos(2T_0)+ \right. \nonumber \\
&& \left. \xi_{n,q}^{(4)}\left(\mu_{1q}(T_2),\nu_{1q}(T_2)\right)\sin(2T_0)\right]\exp(\alpha_{n,q}z)\mathrm{J_0}(\alpha_{n,q}r)\label{eq3.16} \\\nonumber \\
&& \text{and} \nonumber \\\nonumber \\
&&\eta_2(r,T_0,T_2) = \sum_{n = 1}^{\infty}\left[\zeta_{n,q}^{(1)}\cos(\omega_{n,q}T_0) + \zeta_{n,q}^{(3)}(\mu_{1q}(T_2),\nu_{1q}(T_2))\cos(2T_0) + \right. \nonumber \\
&& \left.  \zeta_{n,q}^{(4)}(\mu_{1q}(T_2),\nu_{1q}(T_2))\sin(2T_0)+\zeta_{n,q}^{(5)}(\mu_{1q}(T_2),\nu_{1q}(T_2))\right]\mathrm{J_0}(\alpha_{n,q}r).\label{eq3.17}
\end{eqnarray}

In eq. \ref{eq3.16}, $\xi_{n,q}^{(1)}=0,\xi_{n,q}^{(2)}$ are constants (part of the complementary function for eqn. \ref{eq3.8}a, these being assumed to be constants instead of being functions of $T_2$, as explained earlier). Expressions for  $\xi_{n,q}^{(2)}$, $\xi_{n,q}^{(3)}\left(\mu_{1q}(T_2),\nu_{1q}(T_2)\right)$ and $\xi_{n,q}^{(4)}\left(\mu_{1q}(T_2),\nu_{1q}(T_2)\right)$  are provided in the appendix. Similarly in eq. \ref{eq3.17}, $\zeta_{n,q}^{(1)}$ and $\zeta_{n,q}^{(2)}=0$ are constants while expressions for $\zeta_{n,q}^{(3)}\left(\mu_{1q}(T_2),\nu_{1q}(T_2)\right)$, $\zeta_{n,q}^{(4)}\left(\mu_{1q}(T_2),\nu_{1q}(T_2)\right)$ and $\zeta_{n,q}^{(5)}\left(\mu_{1q}(T_2),\nu_{1q}(T_2)\right)$ are provided in the appendix. Note that these expressions depend on the unknowns $\mu_{1q}(T_2)$ and $\nu_{1q}(T_2)$. Equations for these two unknowns are obtained at $\mathcal{O}(\epsilon^3)$ via elimination of resonant forcing terms for the primary mode.
\subsection{Amplitude equation(s)}
In order to obtain the equations governing $\mu_{1q}(T_2)$ and $\nu_{1q}(T_2)$, it is necessary to carry out the calculation till $\mathcal{O}(\epsilon^3)$ and seek terms which resonante with the primary mode. As the natural frequency of the left hand side of \ref{eq3.18}a is unity for $j=q$ ($\omega_{q,q}^2=1$), we look for terms on the right hand side in \ref{eq3.18}a which are proportional to $\cos(T_0)$ or $\sin(T_0)$. Writing eqn. \ref{eq3.8}a for $i=3$ and $j=q$,
\begin{eqnarray}\label{eq3.18}
	&&\left(D_0^2 + 1\right)p_3^{(q)}(T_0,T_2) = \frac{2}{l_q^2 \mathrm{J_0}^2(l_q)}\int_{0}^{l_q} dr\, r \hspace{1mm} \mathrm{J_0}(r) \left\{D_0\mathscr{N}_3 + \mathscr{M}_3 \right\}. \nonumber \\
\end{eqnarray}
Expressions for $\mathscr{M}_3(r,T_0,T_2),\mathscr{N}_3(r,T_0,T_2)$ necessary to evaluate the right hand of eqn. \ref{eq3.18} are provided in the appendix. Note in particular that $\mathscr{M}_3$ and $\mathscr{N}_3$ both contain terms like $D_2\eta_1$ and $D_2\phi_1$ which lead to time derivatives of $\mu_{1q}(T_2)$ and $\nu_{1q}(T_2)$. Elimination of the coefficients  of $\cos(T_0)$ and $\sin(T_0)$ appearing on the right hand of eqn. \ref{eq3.18} lead us to 
two nonlinear, coupled ordinary differential equations governing evolution of $\mu_{1q}(T_2)$ and $\nu_{1q}(T_2)$ over the slow time scale $T_2$. The algebra is again quite lengthy and we provide only the amplitude equations here. These are of the form
\begin{subequations}\label{eq_ev}
	\begin{align} 
&D_2\mu_{1q}=r_1\mu_{1q}+r_2\nu_{1q}, \;D_2\nu_{1q}=r_3\mu_{1q}-r_1\nu_{1q},\tag{\theequation a,b} \\
& \mu_{1q}(0) = 0,\; \nu_{1q}(0) = -1 \nonumber
	\end{align} 
\end{subequations}
Here $r_{1}$, $r_{2}$ and $r_{3}$ in eqns. \ref{eq_ev}a and \ref{eq_ev}b are nonlinear functions of $\mu_{1q}$ and $\nu_{1q}$ whose expressions are provided in the appendix. Eqns. \ref{eq_ev}a and \ref{eq_ev}b are solved numerically in MATLAB and together with expressons at linear and quadratic order provided earlier in eqns. \ref{eq3.13},\ref{eq3.14},\ref{eq3.16} and \ref{eq3.17}, we obtain a second order solution to the primary CP problem for our initial condition. This analytical prediction is free from any fitting parameters and is tested against numerical simulations of the incompressible Euler's equation with gravity, in the next section.

\subsection{Comparison of theory with numerical simulations}
\begin{figure}
	\centering
	\includegraphics[scale=0.2]{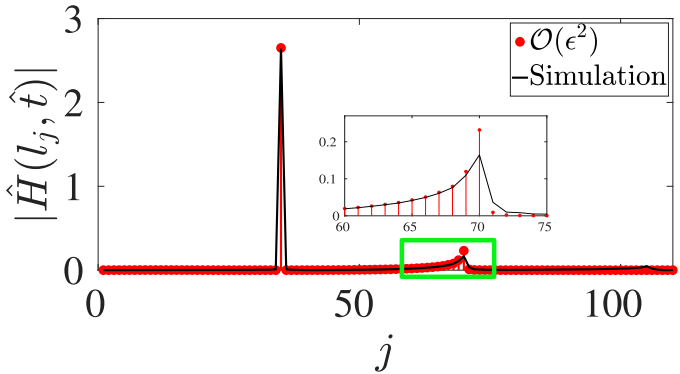}
	\caption{Justification of the neglect of terms in the infinite series beyond $n=70$ i.e. $2q$ in eqns. \ref{eq3.16} and \ref{eq3.17}. The plot presents the Hankel transform of the analytical solution viz. $\epsilon\eta_1 + \epsilon^2\eta_2$ for $\epsilon=0.5$ (Case $2$ in table \ref{tab:kd}) at $\hat{t}=0.483$ s when the interface reaches nearly its maximum height (see fig. \ref{fig_time_evol_eps0.5}, panel (d)). For comparison, the Hankel transform of the interface obtained from numerical simulation of the Euler's equation at this time instant is also presented. It is seen that there is no significant potential energy in modes beyond $n=70$ thereby justifying truncation of the infinite series beyond this $n$. The inset is a blowup of the region highlighted in green and shows gravitational potential energy present in modes with $n \leq 70$.}
	\label{fig_hankel}	
\end{figure}

\begin{figure}
	\centering
	\subfloat[$\hat{t}=0$]{\includegraphics[scale=0.14]{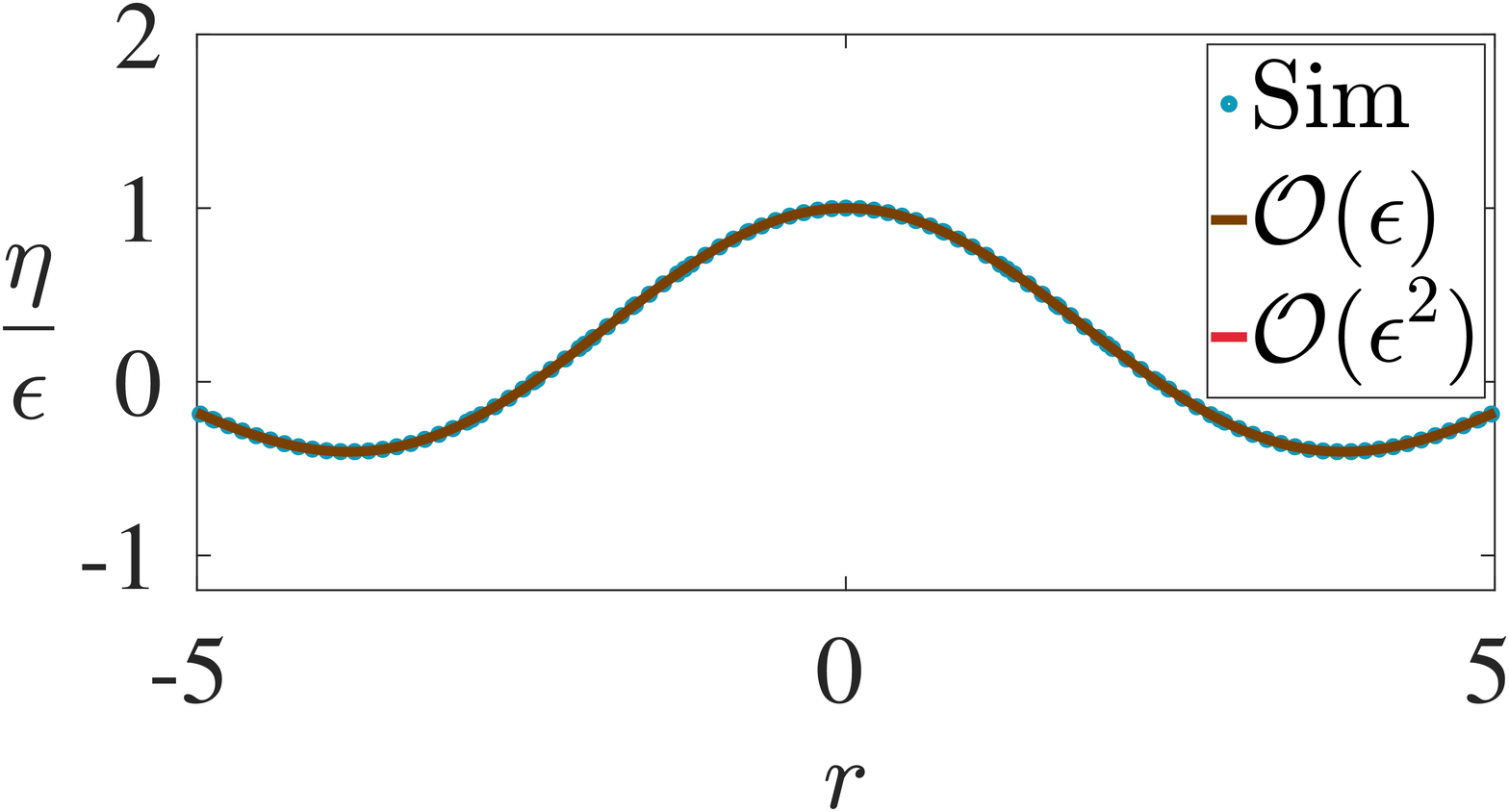}\label{interface1}}\quad
	\subfloat[ $\hat{t}=0.266$ s]{\includegraphics[scale=0.12]{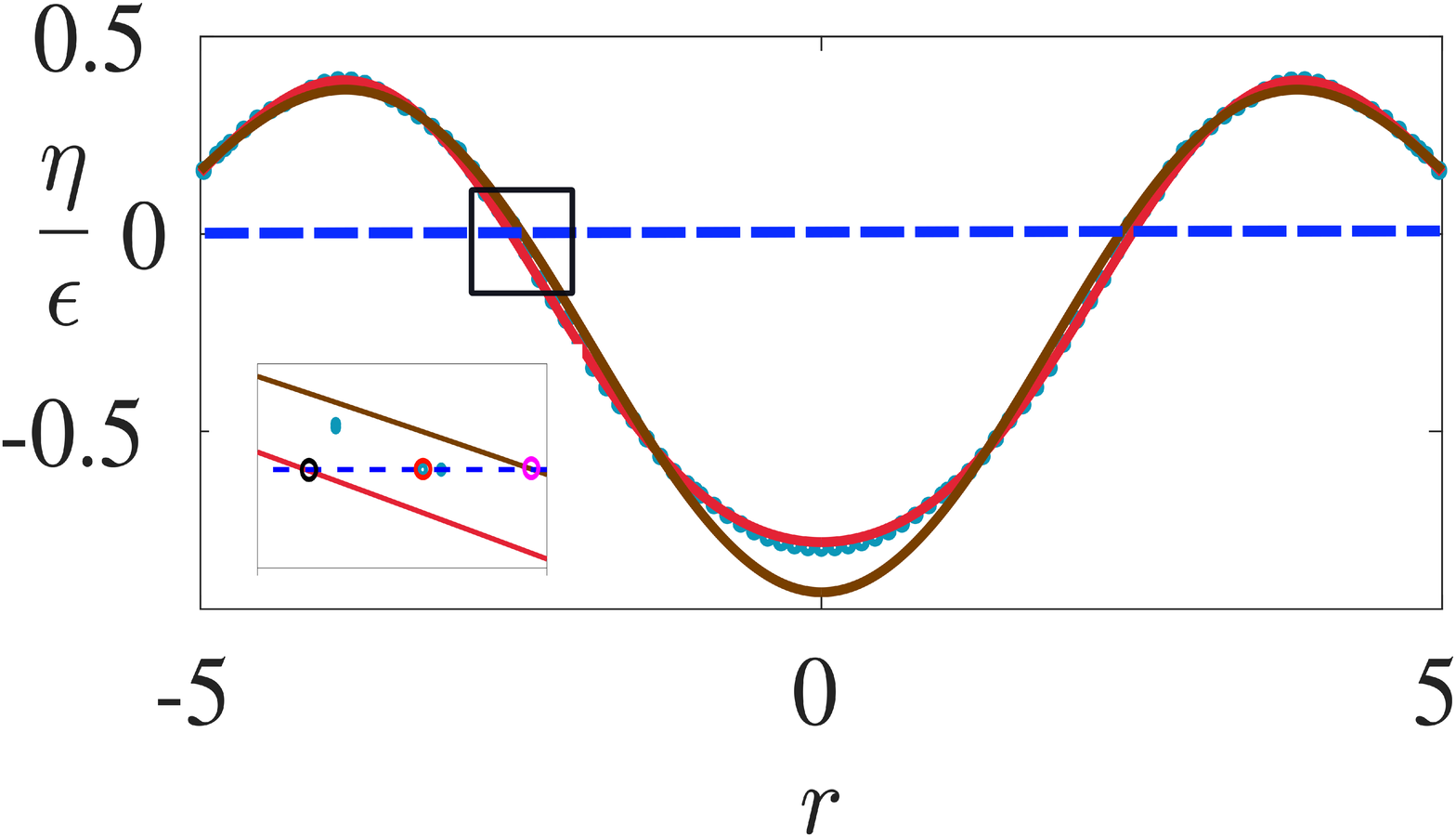}\label{interface2}}\\
	\subfloat[$\hat{t}=0.337$ s]{\includegraphics[scale=0.14]{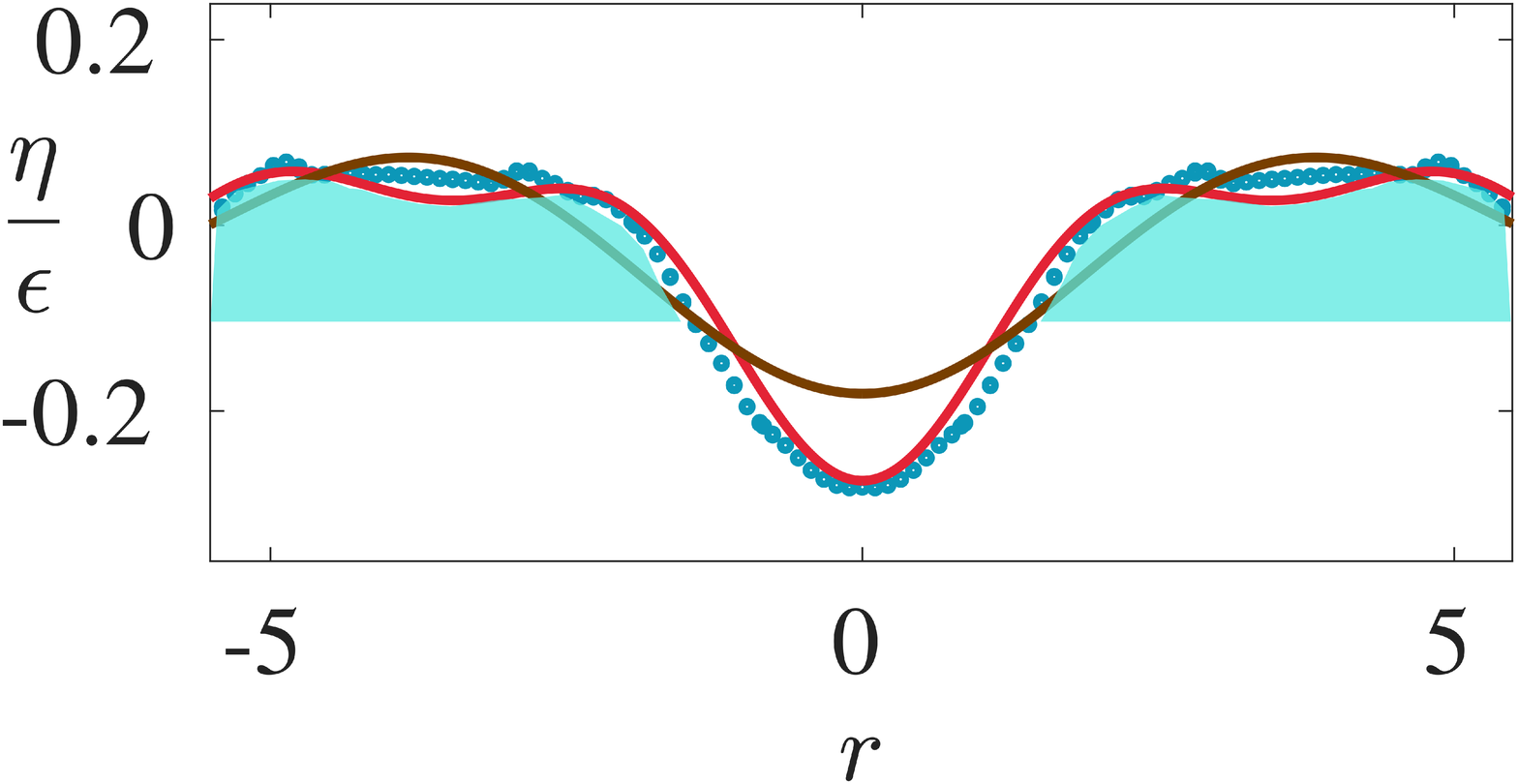}\label{interface3}\quad}
	\subfloat[ $\hat{t}=0.483$ s]{\includegraphics[scale=0.14]{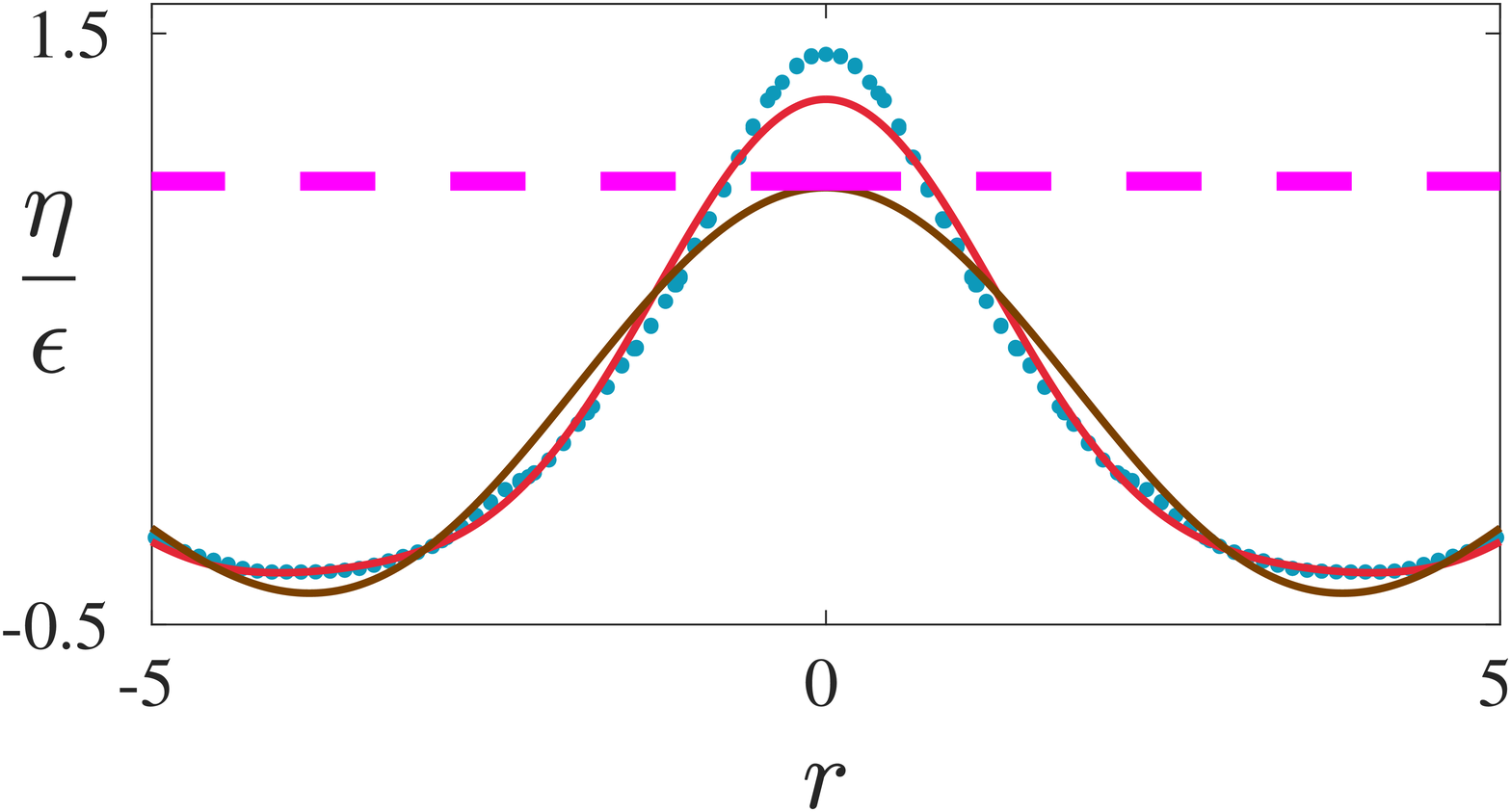}\label{interface4}}
	\caption{Panels (a-d) : Comparison of the interface profile $\frac{\eta}{\epsilon}$ at various instants with linear and weakly nonlinear, analytical solutions. Solid red - $\mathcal{O}(\epsilon^2)$ weakly nonlinear theory, solid brown- $\mathcal{O}(\epsilon)$ linear theory and numerical simulation (Sim, blue dots) for $\epsilon=0.5$ and $l_{35}=110.74$ (Case $2$ in table \ref{tab:kd}). Panel (b) inset depicts a close up view of the first zero crossing of the perturbed interface. Note the qualitative differences between the simulation and linear theory particularly in panels (c) and (d). The shaded region in blue in panel (c) depicts the wave crest whose inward focussing results in an overshoot. Significant overshoot ($\approx 37\%$) is seen in the numerical simulation i.e. above the dashed pink line in panel (d) at $r=0$. Unlike linear theory, the weakly nonlinear theory predicts this overshoot nearly correctly as seen in panel (d).}
	\label{fig_time_evol_eps0.5}	
\end{figure}

In this section, the results from numerical simulations are compared against the theoretical predictions made earlier. In order to do this, we truncate the infinite series in \ref{eq3.16} and \ref{eq3.17} at $n=70$. This number is twice the index of the primary mode, which in the present study has been chosen to be $q=35$. Fig. \ref{fig_hankel} presents the Hankel transform of the interface for $\epsilon=0.5$ from our $\mathcal{O}(\epsilon^2)$ theory, comparing with the Hankel transform of the interface obtained from numerical simulations. This is done at $\hat{t}=0.483$ s and the interface at this instant of time is depicted in fig. \ref{fig_time_evol_eps0.5}, panel (d). It is clearly  seen that at this instant of time, when the interface shows more than $37\%$ overshoot at the symmetry axis (fig. \ref{fig_time_evol_eps0.5}, panel (d)), there is a perceptible amount of potential energy around the second harmonic of the primary mode viz. $q=70$, see inset of figure \ref{fig_hankel}. However, our theoretical model predicts very little potential energy beyond this and consequently, we have truncated our expansion at $n=70$ in expressions \ref{eq3.16} and \ref{eq3.17}.

The numerical simulations have been carried out using the open source code Basilisk \citep{basilisk} employing a cylindrical domain of radius $\hat{R}_0=600$ cm and depth $\hat{H}=300$ cm. As the chosen value of $q=35$, this implies that the (approx.) wavelength of the initial perturbation is far smaller than $\hat{H}=300$, justifying the deep water approximation that we make in our theory. The parameters of various simulations are summarised in table \ref{tab:kd}. In all simulations we have employed free-edge boundary conditions at the wall implying that the interface intersects the solid boundary at $\hat{R}_0$ at an angle of $\pi/2$ at all times.

\begin{table}
	\begin{center}
		\def~{\hphantom{0}}
		\begin{tabular}{ccccccc}
			$\mathrm{Case}$ &  & $\epsilon$   &   $l_q (q=35)$ & $\hat{R}_0$  & $\hat{a}_0$ & $\hat{H}$  \\[3pt]
			1  & & 0.1 & 110.74 & 600 & 0.54 & 300\\
			2  & & 0.5 & 110.74 & 600 & 2.70 & 300\\
			3  & & 0.9 & 110.74 & 600 &4.86  & 300\\
			4   & & 1.3 & 110.74 & 600 & 7.02 &300\\
			5   & & 1.5 & 110.74 & 600 & 8.10 &300\\
			6   & & 1.6 & 110.74 & 600 & 8.64 &300\\
		\end{tabular}
		\caption{Axisymmetric simulations parameters (CGS units) with air (above) - water (below) parameters conducted in Basilisk \citep{basilisk}. All simulations have been conducted with uniform $2048^{2}$ grid.}
		\label{tab:kd}
	\end{center}
\end{table}

\begin{figure}
	\centering
		\includegraphics[scale=0.18]{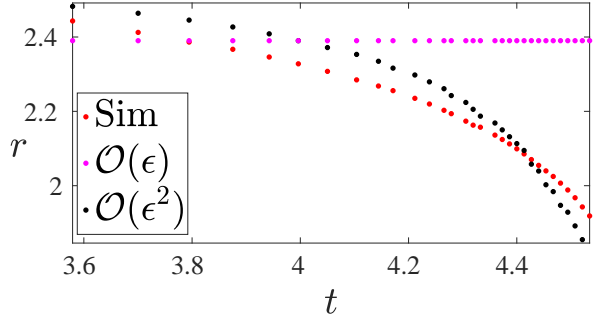}\label{interface5}
			\caption{Radial inward motion of the first zero crossing between panel (b) to panel (c) in fig. \ref{fig_time_evol_eps0.5}. Note from fig. \ref{fig_time_evol_eps0.5}, that within this time window the interface is shaped as a cavity around $r=0$ and the collapse of this cavity occurs nonlinearly, leading to the overshoot seen in fig. \ref{fig_time_evol_eps0.5}. As noted earlier, the inward motion of the first zero crossing of $\eta(r=r^{*},t)$ may be interpreted as radial inward focussing of the two humps shown earlier in fig. \ref{fig_focussing}. As seen, linear theory (pink symbols) does not allow any radial displacement of the zero crossings. However the weakly nonlinear theory is able to describe the inward motion of the nodes correctly qualitatively, although some quantitative differences persist between theory and simulations.}
	\label{fig_rad_inward_nodes}		
\end{figure}

\begin{figure}
	\centering
	\includegraphics[scale=0.2]{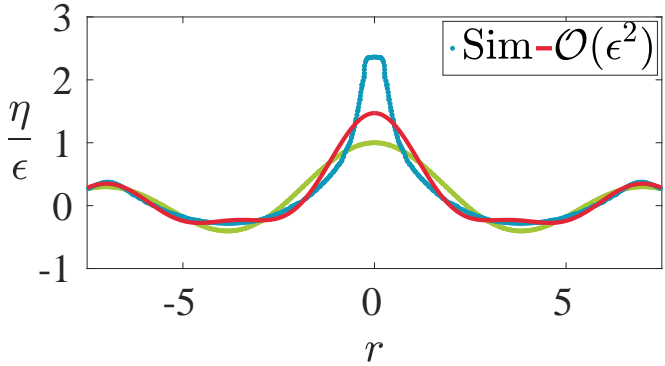}
	\caption{The interface for $\epsilon=0.9, \hat{t}=0.47$ s before the maximum height is reached. The $\mathcal{O}(\epsilon^2)$ theory is unable to describe the slender jet seen in simulations although it manages to get the overshoot correct qualitatively. The line in green represents the initial condition.}
	\label{fig_comp_09}	
\end{figure}

Fig \ref{fig_time_evol_eps0.5} shows the comparison of linear and weakly-nonlinear theory with the simulation for $\epsilon=0.5$. It can be seen that the weakly nonlinear theory captures the simulation profiles quite well. Note in particular panel (d) in this figure, showing clearly that the overshoot seen in the simulation at the symmetry axis over and above unity (pink dashed line in the figure), is predicted well by the $\mathcal{O}(\epsilon^2)$ theory.  This overshoot is the precursor to the slender jet that develops at the axis of symmetry, seen for far larger values of $\epsilon=1.5$, please refer to fig. \ref{fig_surface_eta}. Fig \ref{fig_rad_inward_nodes} shows the inward movement of the first zero crossing of the interface $\eta$ (around $r=2.4$ in fig. \ref{fig_time_evol_eps0.5}, panel b) from panels (b) to (c). Note the good agreement in fig. \ref{fig_rad_inward_nodes} between nonlinear theory and simulation. During this time window, the interface around the axis of symmetry is shaped like a cavity which collapses subsequently, leading to the overshoot seen in panel (d) in fig. \ref{fig_time_evol_eps0.5}. In Fig. \ref{fig_comp_09} we tests the limit of our theory, extending the magnitude of $\epsilon$ to $0.9$ (Case $3$ in table \ref{tab:kd}). It is seen that althought the $\mathcal{O}(\epsilon^2)$ theory shows the overshoot, it significantly underdescribes the jet being unable to capture its slenderness well. For reference, the initial condition is also provided in this figure in green. This value of $\epsilon \sim 0.9$ represents an upper bound on the utility of the $\mathcal{O}\left(\epsilon^2\right)$ theory for describing these jets. While the theory is able to describe the inception (upto $\epsilon =0.5$), it is unsuitable in this parametric regime where nonlinearity becomes very strong. This regime will be analysed in detail in the following section.

We have argued earlier that the development of the jet requires radially inward focussing of surface gravity waves at the axis of symmetry, much akin to the phenomena of bubble collapse albeit at far smaller capillary scales, where radially inward focussing of \textit{capillary waves} are implicated in the formation of a slender jet \citep{duchemin2002jet, gordillo2019capillary}. Predictions from our theoretical model vis-a-vis numerical simulations presented in fig. \ref{fig_time_evol_eps0.5}, demonstrates clearly that such radially inward focussing of \textit{surface gravity waves} at large length-scales, are analogously to be implicated in the generation of similar jets at such length scales. 


\section{Strongly nonlinear regime, inertial regime}
In the previous section, we have studied simulations for $\epsilon < 1$ and the weakly nonlinear theory was able to describe the inception of the jet upto about $\epsilon \approx 0.5$. We now move to the strongly nonlinear regime ($\epsilon >> 1$) where due to strong nonlinearity, the theory becomes inadequate and a slender jet emerges in the numerical simlations (fig. \ref{fig_surface_eta}). Here qualitative understanding may be obtained by referring to the analytical solution on hyperboloidal jets by \cite{longuet1983bubbles} in axisymmetric, extensional flow. It will be seen that this analytical model of \cite{longuet1983bubbles} provides an excellent approximation to the jets that arise in our simulations, when $\epsilon >> 1$. For the benefit of the reader, we briefly summarise the salient points of the theory by \cite{longuet1983bubbles}. 

Consider a time varying hyperboloid whose equation in implicit form is given by eqn. \ref{lh1}. Note that the independent variables in eqn. \ref{lh1} are all dimensional. 
\begin{eqnarray}
f(\hat{x},\hat{y},\hat{z},\hat{t}) = -\frac{\hat{x}^2}{a(\hat{t})^2} - \frac{\hat{y}^2}{b(\hat{t})^2} + \frac{\hat{z}^2}{c(\hat{t})^2} - 1 =0 \label{lh1}
\end{eqnarray} 
\cite{longuet1983bubbles} noted that the incompressible, extensional flow $\hat{\Phi}(\hat{x},\hat{y},\hat{z},\hat{t}) = \frac{1}{2}\left(\left(\frac{\dot{a}}{a}\right)\hat{x}^2 + \left(\frac{\dot{b}}{b}\right)\hat{y}^2 + \left(\frac{\dot{c}}{c}\right)\hat{z}^2\right)$ preserves hyperboloids of the form given by $f(\hat{x},\hat{y},\hat{z},\hat{t})$. In order to determine equations governing $a(\hat{t}), b(\hat{t})$ and $c(\hat{t})$, \cite{longuet1983bubbles} imposed the criteria that $f(\hat{x},\hat{y},\hat{z},\hat{t})$ must also be a material surface of zero pressure thus satisfying $\hat{p} = -\frac{\partial \hat{\Phi}}{\partial \hat{t}} + \frac{1}{2}|\bm{\hat{\nabla}}\hat{\Phi}|^2 + F(\hat{t})=0$ and $\frac{D\hat{p}}{D\hat{t}}=0$. Note in particular the neglect of gravity in the Bernoulli equation which is crucial for the remaining analysis. Alongwith incompressibility, this leads to the following equations for $a(\hat{t}), b(\hat{t})$ and $c(\hat{t})$ viz.
\begin{subequations}\label{lh2}
	\begin{align}
a\ddot{a} = b\ddot{b} = -c\ddot{c} = F(\hat{t}), \quad \frac{\dot{a}}{a} + \frac{\dot{b}}{b} + \frac{\dot{c}}{c}=0, \tag{\theequation a,b}
	\end{align}
\end{subequations}
overdot(s) indicating differentiation. \cite{longuet1983bubbles} showed that equations \ref{lh2}a,b may be integrated to obtain
\begin{eqnarray}
	a(\hat{t})b(\hat{t})c(\hat{t}) = 2L^3,\quad -\dot{a}^2 - \dot{b}^2  + \dot{c}^2 = U^2 \label{lh3}
\end{eqnarray}
where $L$ and $U$ are constants of integration related to initial conditions. For an axisymmetric surface about the $\hat{z}$ axis, we may set $a(\hat{t})=b(\hat{t})$ and for this case one may integrate equations \ref{lh3} to determine $c(\hat{t})$ and in turn $a(\hat{t})$. Interestingly, this exact solution is singular at $\hat{t}=0$ where it predicts a divergent velocity ($\dot{c}\sim \hat{t}^{-1/3}$) and a divergent acceleration ($\ddot{c}\sim \hat{t}^{-4/3}$) at the tip of the hyperboloid viz. at $\hat{z}=c(\hat{t})$ (see expression for $\hat{\Phi}$ earlier). In a time window around this initial `jerky start' to motion, the half-angle $\theta$ between the asymptotes to the hyperbola are predicted to evolve from its initial value $\theta(0) = \tan^{-1}{\sqrt{2}} \approx 54.5^{\circ}$ as \citep{longuet1983bubbles},
\begin{eqnarray}
	\tan\theta(\hat{t}) \sim \sqrt{2}\left[1 + \frac{U}{L}\left(\frac{\hat{t} - \hat{t}_0}{\sqrt{3}}\right)\right]^{-3/2} \label{lh4}
\end{eqnarray}
where $\hat{t}_0$ is time shift that will be necessary for comparison with our simulations, but is zero for \cite{longuet1983bubbles}. Physically, the \cite{longuet1983bubbles} solution implies that a radially converging flow on the $x-y$ plane and an outgoing flow along the $z$ axis (from mass conservation), can lead to very large accelerations at the tip of the free surface, shaped in this case as a hyperboloid at all time. Notably, this is an exact solution to the nonlinear, potential flow equations without gravity and with such a specially shaped free surface.

\begin{figure}
	\centering
	\subfloat[$\epsilon=1.3$]{\includegraphics[scale=0.12]{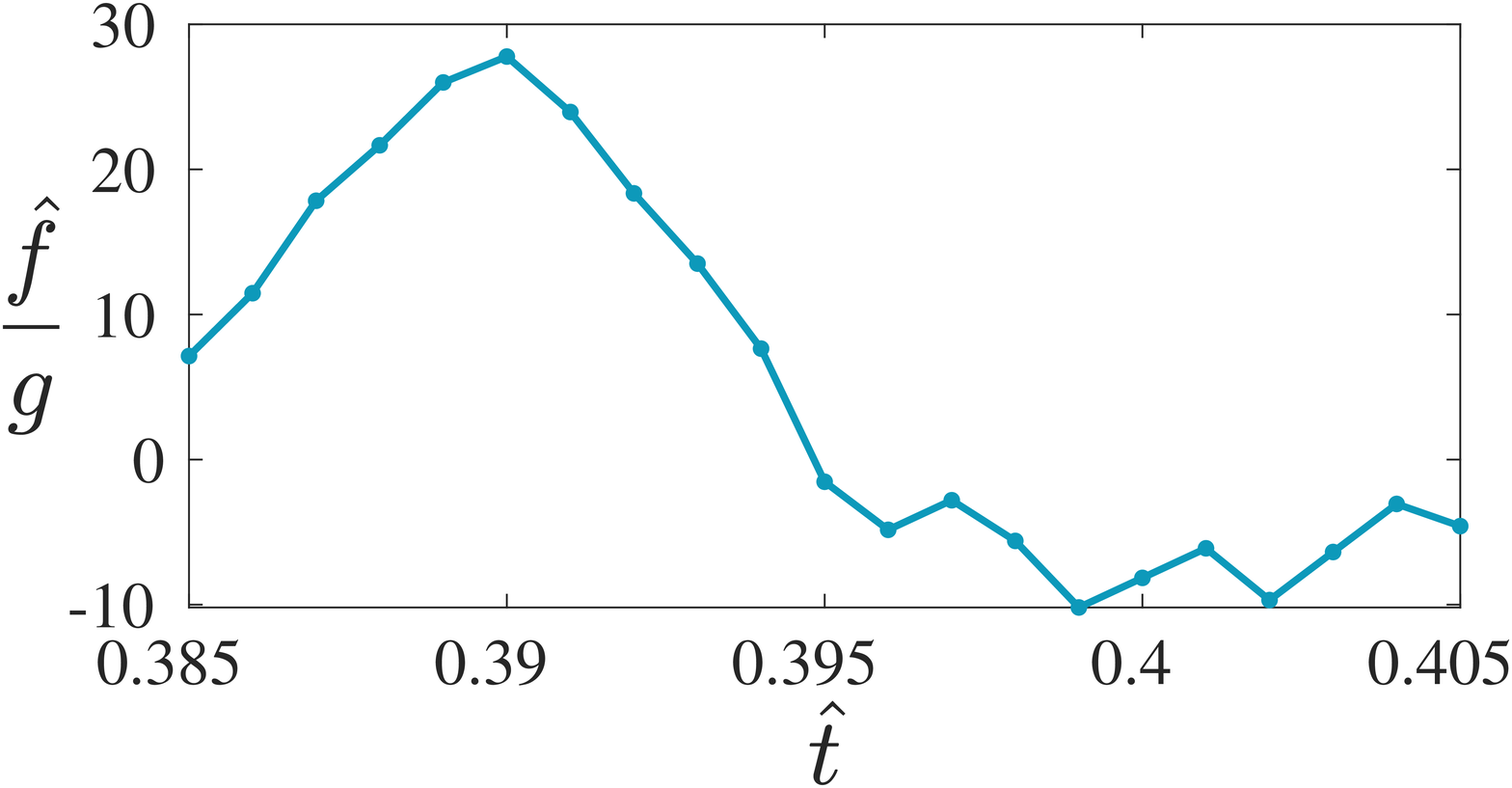}\label{accl1}}\quad
	\subfloat[ $\epsilon=1.3$ ]{\includegraphics[scale=0.12]{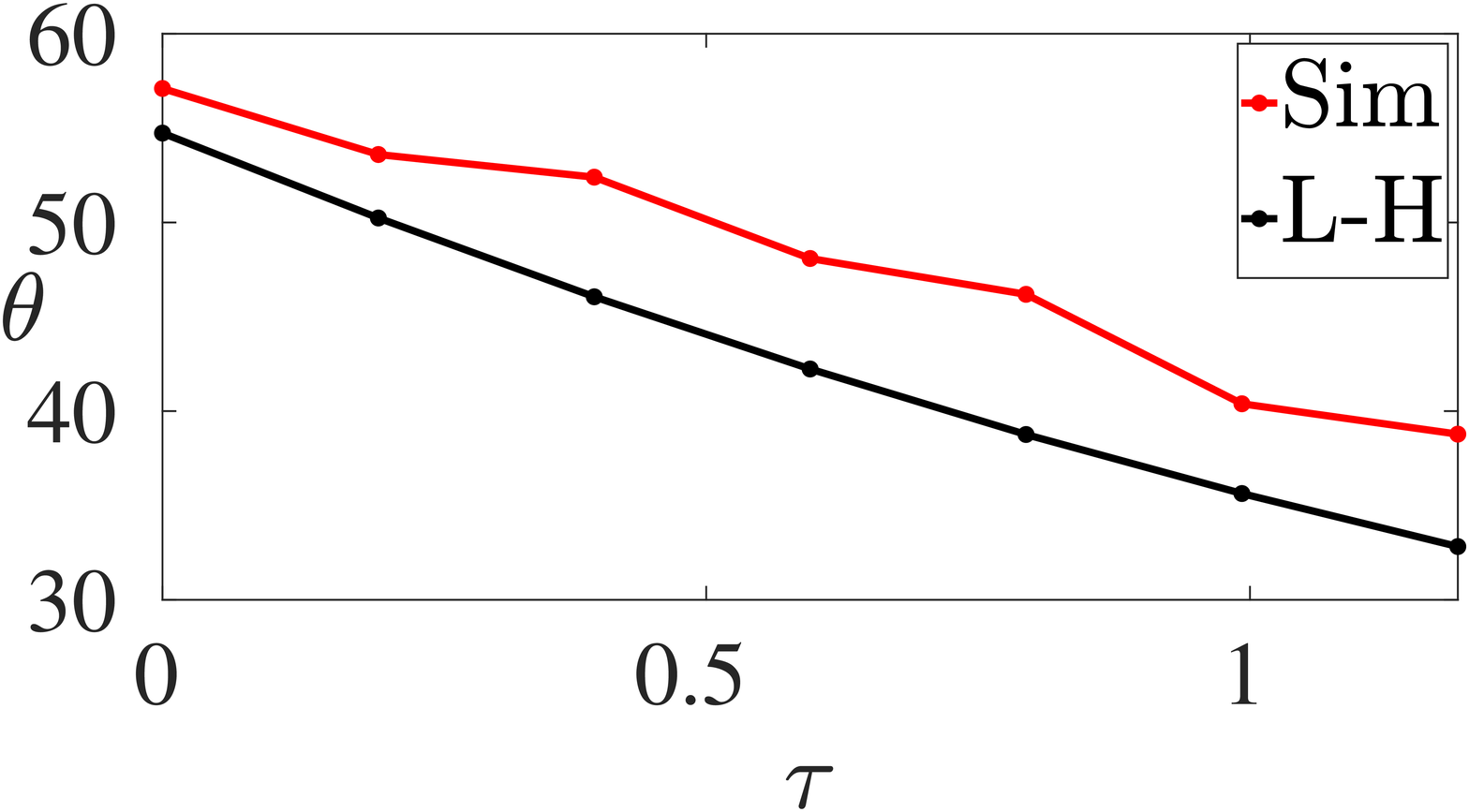}\label{ang1}}\\
	\subfloat[$\epsilon=1.5$]{\includegraphics[scale=0.12]{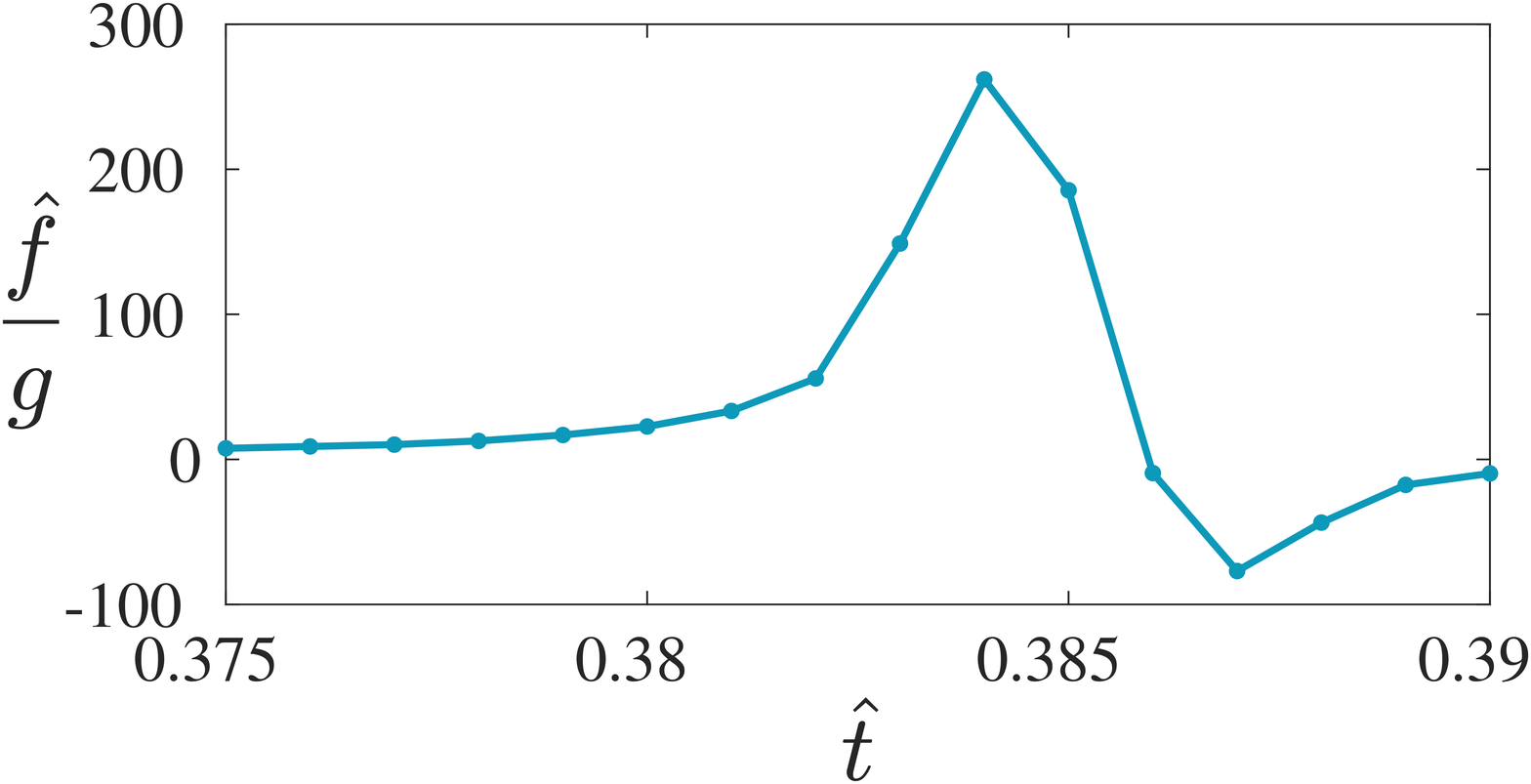}\label{accl2}}\quad
	\subfloat[ $\epsilon=1.5$ ]{\includegraphics[scale=0.12]{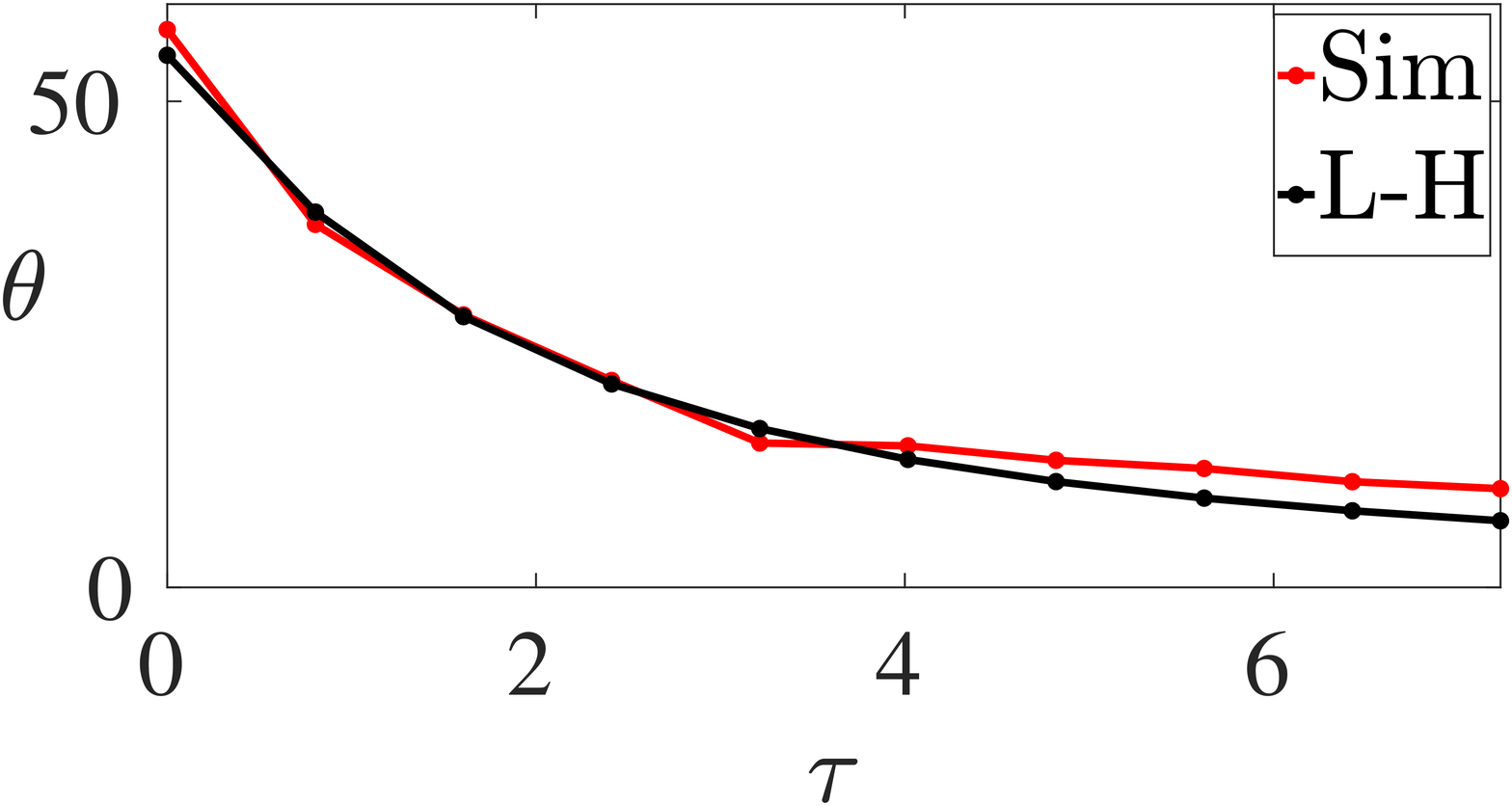}\label{ang2}}\\
	\subfloat[ $\epsilon=1.6$]{\includegraphics[scale=0.12]{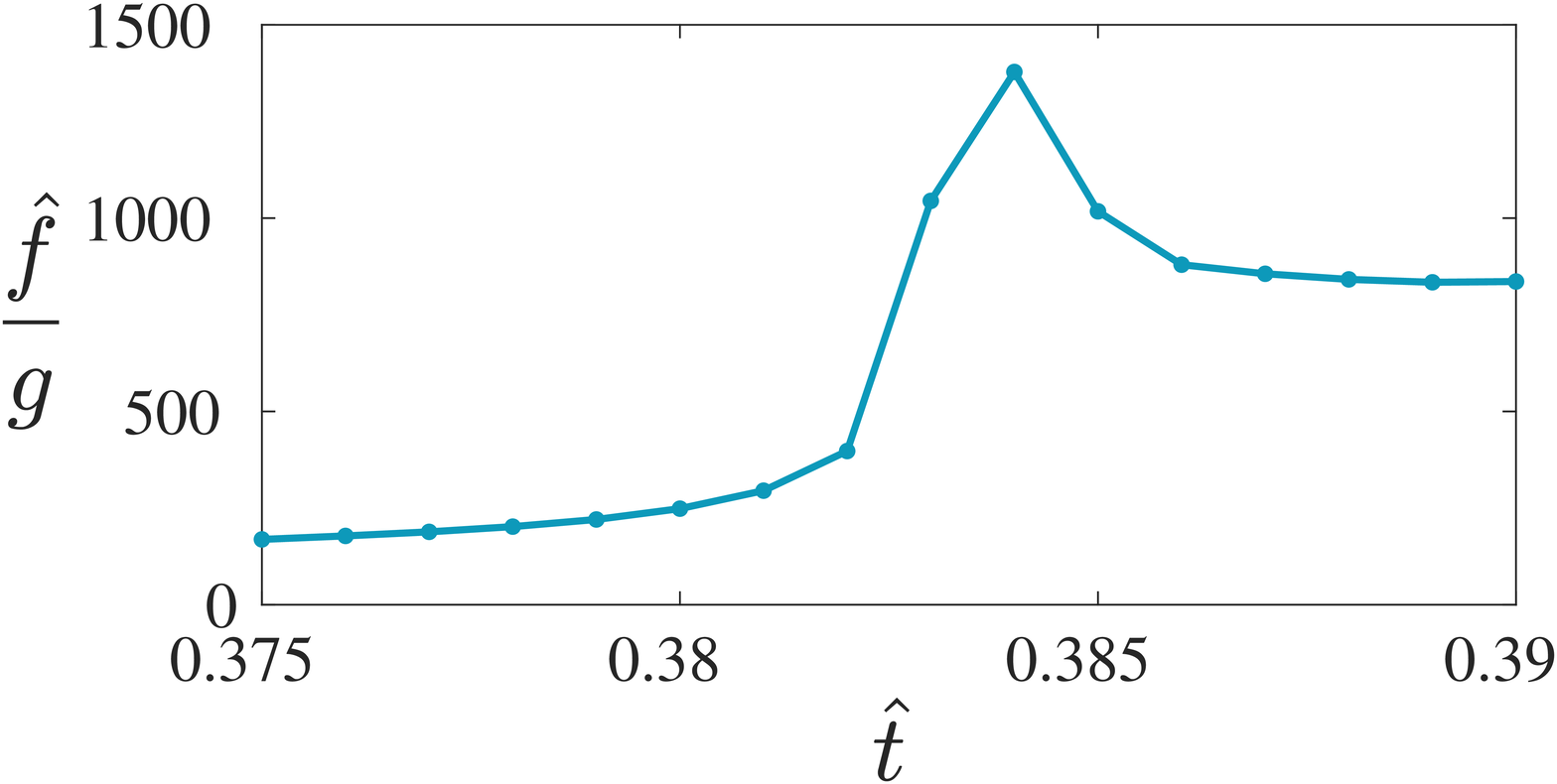}\label{accl3}}\quad
	\subfloat[$\epsilon=1.6$]{\includegraphics[scale=0.12]{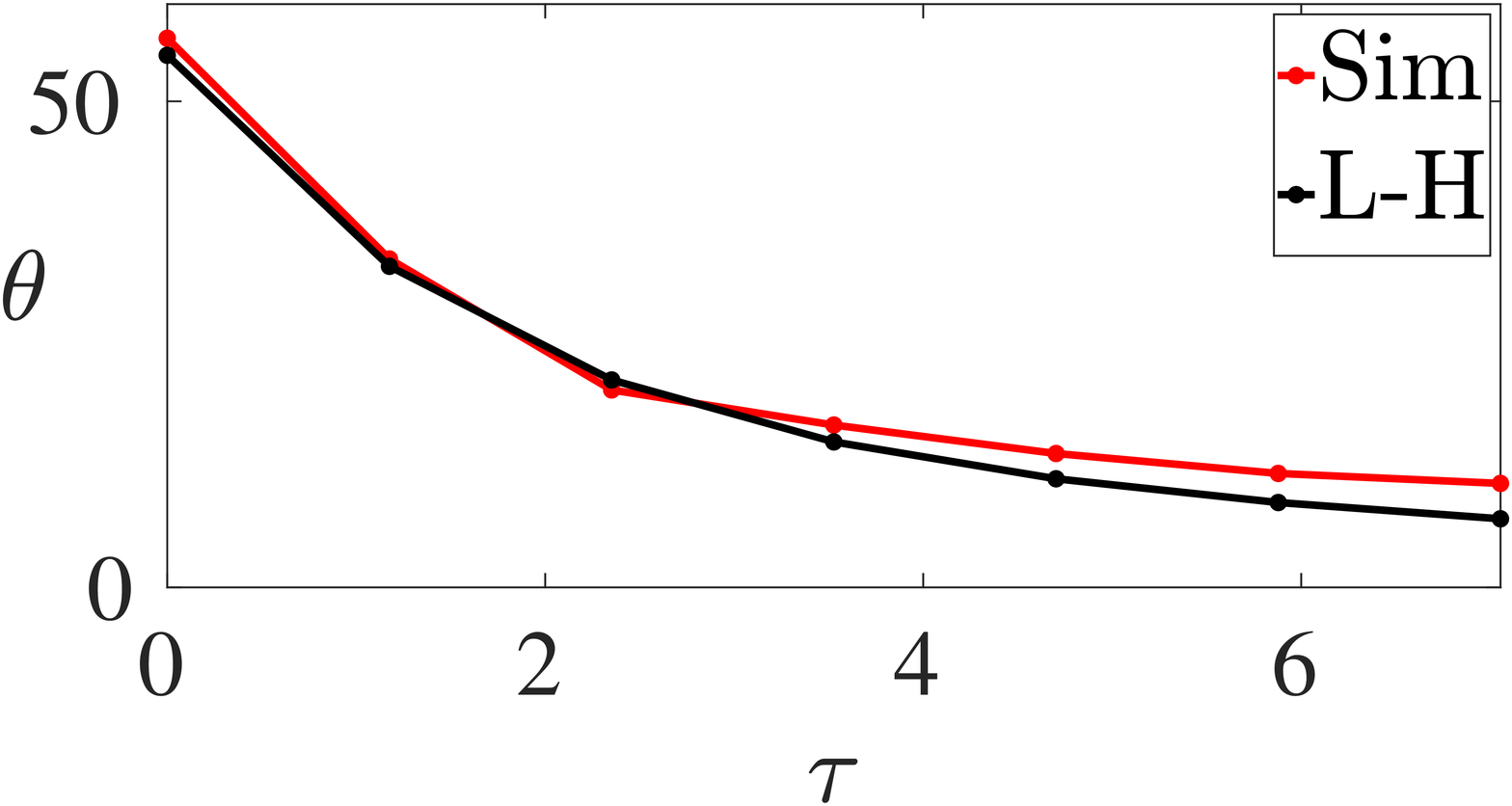}\label{ang3}}
	\caption{Panels (a,c,e) : Acceleration vs time measured from simulations at at various $\epsilon >> 1$. Panels (b,d,f) provide comparison of the angle enclosed at the jet crest starting from $\hat{t}=\hat{t}_0$ with the prediction by \cite{longuet1983bubbles}. The variable $\tau \equiv \frac{U\left(\hat{t} - \hat{t}_0\right)}{L}$ is the non-dimensional time.}
	\label{accl_ang}
\end{figure}

In order to compare our strongly nonlinear jets ($\epsilon >> 1$) with the solution by \cite{longuet1983bubbles}, we first extract the time window where gravity may be safely neglected in our simulations. Figs. \ref{accl1}, \ref{accl2} and \ref{accl3} show the numerically measured acceleration $\hat{f}$ of the interface (in units of gravitational acceleration $g$) at $\hat{r}=0$ for $\epsilon=1.3,\,1.5$ and $1.6$ respectively. The most notable feature is that as $\epsilon >> 1$, the peak acceleration in the jet at the symmetry axis, may exceed $g$ by more than three orders of magnitude, see fig. \ref{accl3}. This implies that gravity may be neglected in our simulations, around a time window centred on this peak, enabling comparison with the \cite{longuet1983bubbles} solution. We have conducted simulations with and without gravity around this time window to verify this conjecture and obtained good agreement in support. In order to compare the angle $\theta$ obtained from our simulations with the prediction from eqn. \ref{lh4}, we set $\hat{t}_0$ to be the instant of time when $\theta$ is closest to the initial angle in the \cite{longuet1983bubbles} model viz. $\approx 54.5^{\circ}$. This definition is also consistent with recent experimental measurements of large scale jets by \cite{mcallister2022wave}, who too compared their jet angle with the \cite{longuet1983bubbles} model. Note that the \cite{longuet1983bubbles} model implies not only an infinite initial acceleration but also infinite initial velocity at the tip of the hyperboloid. We have checked that the velocity of the jet in our simulation ($\epsilon >> 1$) at the instant of such peak acceleration, is orders of magnitude larger than a typical gravity based velocity scale $\sqrt{gL}$, $L$ being the displacement of the interface at $\hat{r}=0$. These justify the relevance of the \cite{longuet1983bubbles} mode for our case and the comparison with this model, that we describe next.  
\begin{figure}
	\centering
	\includegraphics[scale=0.2]{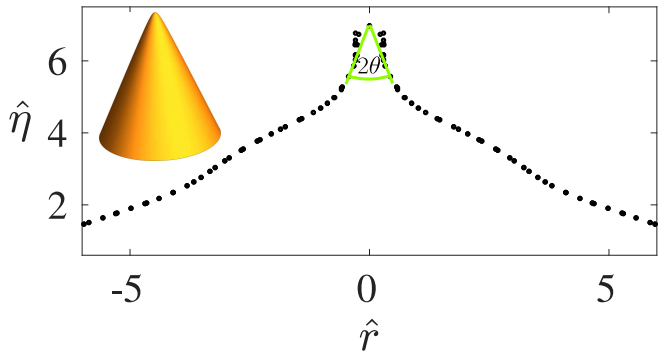}
	\caption{(Inset) A hyperboloid of two sheets symmetric about the vertical $z$ axis with parameters $a=50, b=50$ and $c=200$ in eqn. \ref{lh1} . Only the lower sheet ($z < 0$) is depicted here and serves as a reference for measuring the jet angle $2\theta$, seen in simulations. The main figure depicts the interface $\hat{\eta}$ obtained from numerical simulation with $\epsilon=1.5$, a few time steps after jet inception occurs (i.e. a few time steps after $\hat{t}_0$).}
	\label{fig_hyp}	
\end{figure}

For comparing our simulations with the prediction in eqn. \ref{lh4}, we choose $U$ and $L$ to be equal to the jet velocity and the jet width respectively in the simulation, both values obtained at $\hat{t}=\hat{t}_0$. The angle $\theta$ is measured in our simulations by visually fitting straight lines to the jet tip as depicted in fig. \ref{fig_hyp}. The inset to this figure, shows a hyperboloid of two sheets given in eqn. \ref{lh1} with parameter values indicated in the caption to this figure. Note that the angle $2\theta$ is the angle made by the two straight lines which represent the asymptote to the hyperboloid at large distance. Hence while generating these fits, we have excluded a small region at the jet tip where the slope of the interface is nearly zero. This is consistent with \cite{longuet1983bubbles}, see their figure $3$. Figs. \ref{ang1}, \ref{ang2} and \ref{ang3} present the temporal evolution of the angle $\theta$ with time $\tau \equiv \frac{U\left(\hat{t} - \hat{t}_0\right)}{L}$. The L-H in these figure legends indicates the prediction in eqn. \ref{lh4} by \cite{longuet1983bubbles}. It is clearly seen that for $\epsilon \geq 1.5$, our jets evolve in precisely the manner predicted by eqn. \ref{lh4} the agreement getting better with increasing $\epsilon$.  

\begin{figure}
	\centering
	\subfloat[$\epsilon=1.5$]{\includegraphics[scale=0.13]{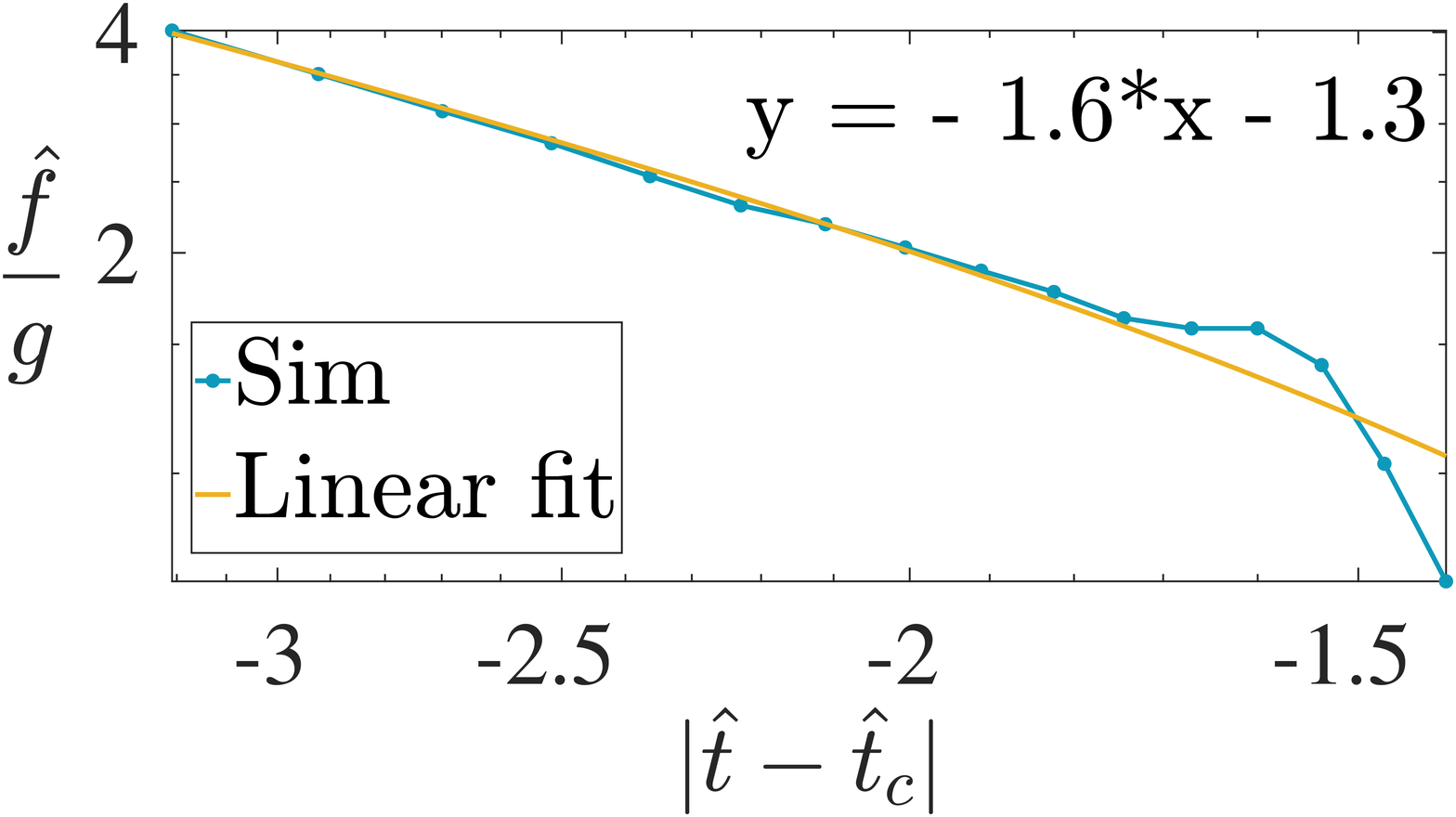}\label{log1}\quad}
	\subfloat[ $\epsilon=1.6$ ]{\includegraphics[scale=0.13]{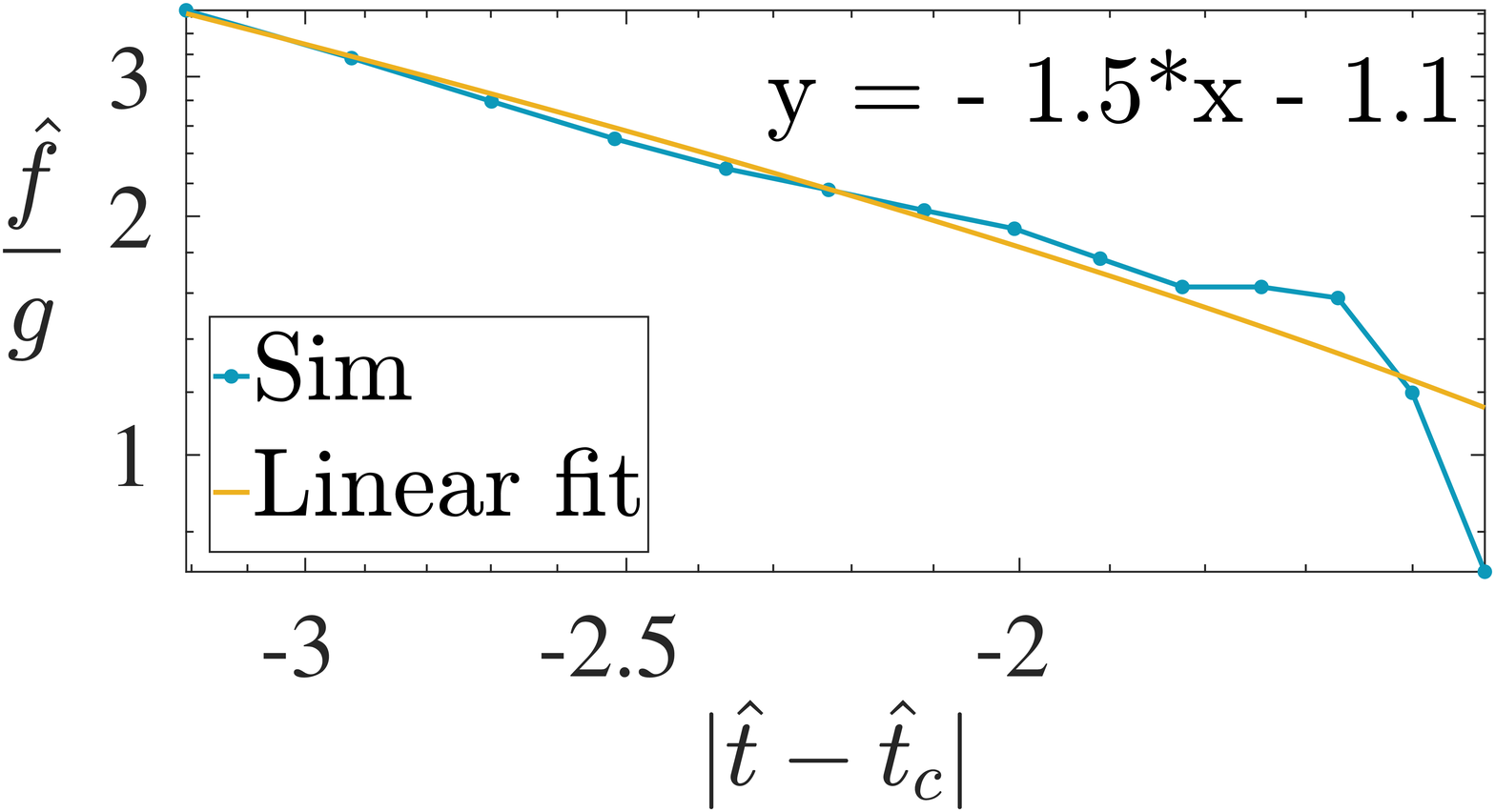}\label{log2}}
	\caption{(a,b) Log-log plot of accelaration vs time. Simulation (blue line) and linear fit (yellow line).}
	\label{log}
\end{figure}

Finally, we also compare the acceleration $\hat{f}$ of the jet (at the symmetry axis) in our numerical simulations with the power law predictions made by \cite{longuet2001vertical_2}. These authors studied high accelerations generated from the collapse of a rounded trough (two dimensional and not axisymmetric unlike our case) reporting accelerations exceeding $100$g. Towards the end of their study, they note that the acceleration of their interface, numerically found to follow a power law of the form $\frac{\hat{f}}{g}\propto |\hat{t}-\hat{t}_c|^{\beta}$ with $\beta=-1.5$ for \cite{longuet2001vertical_2}. We demonstrate in figs. \ref{log1} ($\epsilon=1.5$) and \ref{log2} ($\epsilon=1.6$) that the jets seen in our simulations also display a similar power law viz. $\frac{\hat{f}}{g}\propto |\hat{t}-\hat{t}_c|^{\beta}$, with the exponent being quite close to $-1.5$. Note that $\hat{t}_c$ is defined as the time when the acceleration in our simulations reaches its peak value, as depicted earlier in figs. \ref{accl1}, \ref{accl2} and \ref{accl3}. 
\section{Conclusions}
We have shown in this study, numerically as well as from a first principles theory, that gravity driven focussing of a nonlinear surface wave towards the axis of symmetry can produce jets quite analogous to what also occurs at small scales, for pure capillary waves. Our theory developed using multiple scale analysis, leads to novel amplitude equations governing the modulation of the primary mode, excited initially. The solution to these equations are found to be able to describe the formation of the jet upto $\epsilon \sim 0.5$, capturing the overshoot qualitatively, albeit quantitative differences with numerical simulations are also detected. As $\epsilon$ is increased to nearly $\mathcal{O}(1)$, the weakly non-linear theory becomes expectedly inaccurate while simulations in this regime, show slender jets shooting up with intense accelerations. Here, we find agreement of the time evolution of these jets around the symmetry axis, with a self-similar solution due to \cite{longuet1983bubbles}. Remarkably this solution was derived by \cite{longuet1983bubbles} for the purely inertial regime with zero gravity and we find clear evidence of such a time window in our simulations. In marked contrast to the pure surface tension driven jets at small scales studied recently in \cite{kayal2022dimples}, where surface tension is a part of the self-similar evolution via Keller-Miksis scales \citep{keller1983surface}, we find here that in the pure gravity driven case and for $\epsilon >> 1$, there is a time window when gravitational acceleration at the jet tip becomes negligible compared to the pressure gradient driven, inertial acceleration of the fluid. During this, the interface around the jet tip evolves self-similarly showing striking agreement with the exact solution provided by \cite{longuet1983bubbles}. Our study discusses reasons for why such an agreement may be expected and also explains the physical mechanism for jet formation based on mass conservation, thus independent of dynamics of the problem. These in turn, provides insights into why these jets may be generically observed accompanying cavity collapse phenomena across scales, spanning several orders of magnitude \citep{lee2011size,mcallister2022wave}. 

In conclusion, we note a remark by \cite{longuet2001vertical_1} (page $496$, first paragraph) who remarked that in axisymmetric geometry, the flow focussing responsible for jet formation, would be far more intense. We think that the extra-ordinarily high accelerations ($\approx 10^3g$) seen in our simulations for $\epsilon >> 1$, is due to the radial geometry which induces a $1/r$ divergence as the symmetry axis is approached. Thus the peak accelerations in our simulations are about $10$ times more than that of \cite{longuet2001vertical_2} (two dimensional) where peak accelerations upto $100$ g are reported. Finally we note that the theory developed here, may be viewed as a weakly nonlinear solution to the primary CP problem in axisymmetric coordinates, which provides insights into development of sharply shooting jets in a radial confined geometry. Our study has several implications for such large scale jets generated, for example, in the ocean as well as in other geophysical contexts such as impact craters \citep{lherm2022rayleigh}.
\section*{Declaration of interests}
The authors report no conflict of interest.
\section*{Acknowledgements:} Financial support from DST-SERB (Govt. of India) grants  EMR/2016/000830, CRG/2020/003707 and MTR/2019/001240 on research topics related to waves, jet formation, cavity collapse, bubble bursting and the Cauchy-Poisson problem
are gratefully acknowledged. The Ph.D. tenure of LK is supported by the Prime Minister's Research Fellowship (PMRF), Govt. of India and is gratefully acknowledged. We thank Dr. Vatsal Sanjay, Univ. Twente, Netherlands for sharing with us his code. This was used to generate the initial condition in fig. \ref{fig_bubble} for the Basilisk simulation.

\section*{Appendix: Analytical expressions}
The expressions for $\mathscr{M}_2, \mathscr{N}_2, \mathscr{M}_3, \mathscr{N}_3$ are
\footnotesize
\begin{eqnarray}
&& \mathscr{M}_2(r,T_0,T_2) \equiv \left(\frac{\partial \phi_1}{\partial r}\right)\left(\frac{\partial \eta_1}{\partial r} \right) - \eta_1\left(\frac{\partial^2\phi_1}{\partial z^2}\right), \quad z = 0 \nonumber\\
&& \mathscr{N}_2(r,T_0,T_2) \equiv -\eta_1\left(\frac{\partial^2 \phi_1}{\partial z \partial T_0}\right) -\frac{1}{2}\left\{\left(\frac{\partial\phi_1}{\partial r}\right)^2+\left(\frac{\partial\phi_1}{\partial z}\right)^2\right\}, \quad z = 0\nonumber\\
&&\mathscr{M}_3(r,T_0,T_2) \equiv \left( \frac{\partial \phi_1}{\partial r} \right)\left( \frac{\partial \eta_2}{\partial r} \right) + \left( \frac{\partial \phi_2}{\partial r} \right)\left( \frac{\partial \eta_1}{\partial r} \right) + \eta_1\left(\frac{\partial^2\phi_1}{\partial z \partial r}\right)\left( \frac{\partial \eta_1}{\partial r}\right) - \eta_1\left(\frac{\partial^2 \phi_2}{\partial z^2} \right) \nonumber\\
&& - \eta_2\left(\frac{\partial^2 \phi_1}{\partial z^2} \right) - \frac{\eta_1^2}{2}\left( \frac{\partial^3 \phi_1}{\partial z^3} \right) + \left( \frac{\partial \eta_1}{\partial T_2}\right), \quad z = 0\nonumber\\
&&\mathscr{N}_3(r,T_0,T_2) \equiv -\eta_2\left(\frac{\partial^2\phi_1}{\partial z \partial T_0} \right) - \eta_1\left(\frac{\partial^2\phi_2}{\partial z \partial T_0} \right) - \frac{\eta_1^2}{2}\left(\frac{\partial ^3 \phi_1}{\partial z^2 \partial T_0}  \right) - \frac{\eta_1}{2}\frac{\partial}{\partial z}\left\{\left(\frac{\partial\phi_1}{\partial r}\right)^2+\left(\frac{\partial\phi_1}{\partial z}\right)^2\right\} \nonumber\\
&& - \left( \frac{\partial \phi_1}{\partial r} \right)\left( \frac{\partial \phi_2}{\partial r} \right) - \left( \frac{\partial \phi_1}{\partial z} \right)\left( \frac{\partial \phi_2}{\partial z} \right)- \left( \frac{\partial \phi_1}{\partial T_2}\right), \quad z = 0 \nonumber \\\nonumber\\\nonumber\\
&& \xi_{j,q}^{(1)}\equiv 0, \quad  \xi_{j,q}^{(2)}\equiv\frac{2}{\mathrm{J_0}^2(l_j)\omega_{j,q}\left(\omega_{j,q}^2-4\right)}\left[(\alpha_{j,q}-2)I_{0-q,0-q,0-j}+2I_{1-q,1-q,0-j}\right]\nonumber\\
&& \xi_{j,q}^{(3)}(\mu_{1q}(T_2),\nu_{1q}(T_2))\equiv-\frac{4\mu_{1q}(T_2)\nu_{1q}(T_2)}{\mathrm{J_0}^2(l_j)\left(\omega_{j,q}^2-4\right)}\left[I_{0-q,0-q,0-j}+I_{1-q,1-q,0-j}\right]\nonumber\\
&& \xi_{j,q}^{(4)}(\mu_{1q}(T_2),\nu_{1q}(T_2))\equiv\frac{2\left(\mu_{1q}^2-\nu_{1q}^2\right)}{\mathrm{J_0}^2(l_j)\left(\omega_{j,q}^2-4\right)}\left[I_{0-q,0-q,0-j}+I_{1-q,1-q,0-j}\right]\nonumber \nonumber \\\nonumber\\\nonumber\\
&&  \zeta_{j,q}^{(1)}\equiv -\frac{2}{\mathrm{J_0}^2(l_j)\left(\omega_{j,q}^2-4\right)}\left[(\alpha_{j,q}-2)I_{0-q,0-q,0-j}+2I_{1-q,1-q,0-j}\right]\nonumber\\
&& \zeta_{j,q}^{(2)}\equiv0, \quad  \zeta_{j,q}^{(3)}(\mu_{1q}(T_2),\nu_{1q}(T_2))\equiv\frac{\nu_{1q}^2-\mu_{1q}^2}{2\mathrm{J_0}^2(l_j)\left(\omega_{j,q}^2-4\right)}\left[\left(3\alpha_{j,q}-4\right)I_{0-q,0-q,0-j}+\left(\alpha_{j,q}+4\right)I_{1-q,1-q,0-j}\right]\nonumber\\
&& \zeta_{j,q}^{(4)}(\mu_{1q}(T_2),\nu_{1q}(T_2))\equiv-\frac{\mu_{1q}(T_2)\nu_{1q}(T_2)}{\mathrm{J_0}^2(l_j)\left(\omega_{j,q}^2-4\right)}\left[\left(3\alpha_{j,q}-4\right)I_{0-q,0-q,0-j}+\left(\alpha_{j,q}+4\right)I_{1-q,1-q,0-j}\right]\nonumber\\
&& \zeta_{j,q}^{(5)}(\mu_{1q}(T_2),\nu_{1q}(T_2))\equiv\frac{1}{2\mathrm{J_0}^2(l_j)}\left(\mu_{1q}^2+\nu_{1q}^2\right)\left(I_{0-q,0-q,0-j}-I_{1-q,1-q,0-j}\right)\nonumber
\end{eqnarray}
\begin{eqnarray}
\footnotesize
&& r_1\equiv-\frac{1}{\mathrm{J_0}^2(l_q)}\left[ \sum_{m = 1}^{\infty}\left\{\left(-\zeta^{(4)}_{m,q}(\mu_{1q}(T_2),\nu_{1q}(T_2))+ \frac{1}{2}\left(\alpha_{m,q}^2-\alpha_{m,q}\right)\xi^{(3)}_{m,q}(\mu_{1q}(T_2),\nu_{1q}(T_2))\right)\mathcal{I}_{0-q,0-m,0-j}\right.\right.\nonumber\\     &&\left. + \frac{1}{2}\alpha_{m,q}\zeta^{(5)}_{m,q}(\mu_{1q}(T_2),\nu_{1q}(T_2)) \mathcal{I}_{1-q,1-m,0-j}     \right\}\Bigg]\nonumber\\
&& r_2\equiv-\frac{1}{\mathrm{J_0}^2(l_q)}\left[ \sum_{m = 1}^{\infty}\left\{\left(\zeta^{(3)}_{m,q}(\mu_{1q}(T_2),\nu_{1q}(T_2))+ \frac{1}{2}\left(\alpha_{m,q}^2-\alpha_{m,q}\right)\xi^{(4)}_{m,q}(\mu_{1q}(T_2),\nu_{1q}(T_2))\right)\mathcal{I}_{0-q,0-m,0-j}\right.\right.\nonumber\\     &&\left. + \alpha_{m,q}\bigg(\zeta^{(5)}_{m,q}(\mu_{1q}(T_2),\nu_{1q}(T_2))-\frac{1}{2}\zeta^{(3)}_{m,q}(\mu_{1q}(T_2),\nu_{1q}(T_2))\bigg) \mathcal{I}_{1-q,1-m,0-j}     \right\}\Bigg]\nonumber\\
&& r_3\equiv\frac{1}{\mathrm{J_0}^2(l_q)}\left[ \sum_{m = 1}^{\infty}\left\{\left(-\zeta^{(3)}_{m,q}(\mu_{1q}(T_2),\nu_{1q}(T_2))- \frac{1}{2}\left(\alpha_{m,q}^2-\alpha_{m,q}\right)\xi^{(4)}_{m,q}(\mu_{1q}(T_2),\nu_{1q}(T_2))\right)\mathcal{I}_{0-q,0-m,0-j}\right.\right.\nonumber\\
&&\left. + \alpha_{m,q}\bigg(\zeta^{(5)}_{m,q}(\mu_{1q}(T_2),\nu_{1q}(T_2))+\frac{1}{2}\zeta^{(3)}_{m,q}(\mu_{1q}(T_2),\nu_{1q}(T_2))\bigg) \mathcal{I}_{1-q,1-m,0-j}     \right\}\Bigg]  \nonumber
\end{eqnarray}

\bibliographystyle{jfm}%
\bibliography{jfm}
\end{document}